\documentclass{aa}  

\usepackage{graphicx}
\usepackage{txfonts}
\usepackage{array}
\usepackage{placeins}

\begin{document} 
\setlength{\extrarowheight}{4pt}

   \title{The MeerKAT Fornax Survey}
   \subtitle{VII. Characterisation of the Fornax cluster's magnetic field and new insights on magnetisation in large scale systems}
   \titlerunning{The Fornax cluster's magnetic field with the MeerKAT Fornax Survey}

   \author{F. Loi\inst{1},
          M. Murgia\inst{1},
          F. Govoni\inst{1},
          P. Serra\inst{1},
          V. Vacca\inst{1}, 
          F. M. Maccagni\inst{1,3},
          P. Kamphuis\inst{2},
          and
          D. Kleiner\inst{4,1} 
          }
    \authorrunning{F. Loi et al.}
   \institute{INAF - Osservatorio Astronomico di Cagliari,
              via della scienza 5, Selargius, Italy\\
              \email{francesca.loi@inaf.it}
         \and
              Ruhr University Bochum, Faculty of Physics and Astronomy, Astronomical Institute (AIRUB), 44780 Bochum, Germany
        \and
            Wits Centre for Astrophysics, School of Physics, University of the Witwatersrand, 1 Jan Smuts Avenue, 2000, Johannesburg, South Africa
        \and
            Netherlands Institute for Radio Astronomy (ASTRON), Oude Hoogeveensedijk 4, 7991 PD Dwingeloo, The Netherlands
             }

   \date{Received March 20, 2026; accepted April 07, 2026}

  \abstract
  {Large scale magnetic fields in galaxy clusters can influence their physics and the evolution of the cluster-embedded galaxies. These properties remain poorly constrained due to a historical lack of high-sensitivity and high-resolution spectro-polarimetric data. Thanks to the advent of the Square Kilometre Array pathfinders and precursors this is now dramatically changing.
  By exploiting the densest rotation measure (RM) grid produced to date with broadband spectro-polarimetric data in the context of the MeerKAT Fornax Survey and presented in a previous paper (508 sources over 6.35\,deg$^2$), we aim to study in detail the Fornax cluster's magnetic field.
  We compare the RM grid properties with numerical simulations to constrain the strength and the structure of the intra-cluster magnetic field.
  We model the magnetic field power spectrum with a power--law and we find a slope of 2.7$^{+0.2}_{-0.4}$ fluctuating between a minimum and a maximum scale of 1.01$^{+0.01}_{-0.02}$ and 15$^{+9}_{-2}$\,kpc respectively. It has a central strength of 5.0$^{+0.3}_{-0.4}$\,$\muup$G decreasing with the thermal plasma density according to a power--law exponent $\eta$=1.6$^{+0.3}_{-0.5}$, the highest value to date in large scale systems. By analysing a sample of 17 galaxy clusters and groups with magnetic field estimates from the literature, we observe larger auto-correlation lengths in the case of massive merging clusters and lower values for relaxed clusters and low-mass clusters/galaxy groups. We also observe a systematic increase of the central magnetic field strength as a function of central density, B$_0\propto$\,n$_0^{(0.38\pm0.14)}$.
  Finally, we argue that the steepening of the Fornax cluster's magnetic field profile and its relatively high central strength could be indicative of a recent re-amplification at the centre due to the extended central radio galaxy. The sample analysis supports the proposed scenario, although more detailed magnetic field studies conducted using consistent modelling on larger samples are needed to better understand magnetisation in clusters and groups.}


   \maketitle

\section{Introduction}
Cosmic magnetism is one of the scientific drivers of the Square Kilometre Array telescope. Indeed, the origin and the evolution of large scale magnetic fields embedded in galaxy clusters and along the cosmic web filaments is still uncertain.\\
Large scale magnetic fields in clusters can affect galaxy evolution, as recently observed by \cite{muller2021}, and accelerate particles creating the so-called diffuse radio sources \citep[see the review by][]{vanweeren2019}. Moreover, they can affect the galaxy cluster physics altering the heat conduction, the gas mixing in the intra-cluster medium (ICM), and the propagation of cosmic rays \citep[e.g.][]{Pfrommer2017AIPC.1792c0003P,Ruszkowski2010ApJ...713.1332R}. Achieving a comprehensive understanding of galaxy cluster physics requires constraining the properties of the intra-cluster magnetic fields.\\
This objective can be achieved by utilizing both background and cluster-embedded polarized sources to reconstruct the so-called rotation measure (RM). This is made possible by the Faraday effect, whereby 
the polarisation plane of a background source crossing an external magneto-ionic medium (the galaxy cluster) rotates as a function of the RM and of the rest frame wavelength squared $\lambda^2$:
\begin{equation}
\rm    \Delta\Psi = RM \cdot \lambda^2,
\end{equation}
where $\Delta\Psi$ is the change in polarisation angle.
The RM is linked to the properties of the magnetic field through:
\begin{equation}
\rm    RM = 812 \int_0^L B_{||} \cdot n_e \cdot dl,
\label{eq:rm}
\end{equation}
where n$_e$ is the thermal plasma density in cm$^{-3}$, B$_{||}$ is line-of-sight parallel component of the intervening magnetic field in $\muup$G, L is the distance between the observer and the source in kpc, and the RM is in units of rad m$^{-2}$. \\
The RM properties reflect those of the intervening magnetic field and thermal plasma. The RM values associated with a galaxy cluster are in general fluctuating on a range of scales reflecting its turbulent $\muup$G-level magnetic field. They show a radial profile with a maximum at the cluster centre decreasing outwards. These profiles can be modelled with analytical approaches \cite{Lawler1982ApJ...252...81L,Felten1996ASPC...88..271F} to derive an estimate of the intra-cluster magnetic field \citep[e.g][]{Feretti1999A&A...344..472F}.
Numerical tools can be used to improve the modelling and to measure the intra-cluster magnetic fields with good accuracy from the observed RM \citep[e.g.][]{ensslin2003A&A...401..835E,vogt2003A&A...412..373V,murgia2004}.
However, the use of advanced numerical tools to estimate large scale magnetic fields from the RM measurements is not trivial and can require significant computational effort. An important limitation comes from the finite sensitivity and resolution of the radio telescopes which are able to detect a limited number of polarized sources per square degree in a limited number of galaxy clusters, namely 20 hitherto at most \citep{vogt2003A&A...412..373V, murgia2004, govoni2006, guidetti2008, laing2008, bonafede2010, guidetti2010, vacca2010A&A...514A..71V, vacca2012, govoni2017, stuardi2021, Vacca2022MNRAS.514.4969V, Pagliotta2025A&A...700A.139P, derubeis2024A&A...691A..23D, khadir2025arXiv251118532K, alonso2026A&A...705A.143A}. To better understand how intra-cluster magnetic fields evolve in different environments, under which mechanisms, and the impact on the physics of cluster galaxies, it is of paramount importance to increase the sample.

In a previous work \citep[hereafter L25,][]{Loi2025A&A...694A.125L} we presented the broadband spectro-polarimetric data acquired within the MeerKAT Fornax Survey \citep[MFS,][]{serra2023}. This is a key survey of the MeerKAT telescope which started observing in 2020. It is covering the Fornax cluster and the infalling Fornax A group with 91 pointings. 
With the broadband data we reconstructed the densest RM grid ever constructed, $\sim$80 polarized sources per square degree (L25). The main results of L25 are summarized in Sect. 2. \\
In this paper, we will use the RM grid to constrain the properties of the Fornax cluster's magnetic field. Studying the intra-cluster magnetic field of Fornax is an important step in the investigation of large scale magnetic fields in clusters for two main reasons: first, the cluster is a low-mass cluster having a virial mass of M$_{\rm vir}$ = 5 $\times$ 10$^{13}$ M$_{\odot}$ \citep{Drinkwater2001ApJ...548L.139D} and a core radius of $\sim$173\,kpc \citep{Reiprich2002ApJ...567..716R}, therefore it is
representative of the majority of clusters in the Universe; second, it is nearby, at a distance of $\sim$20\,Mpc \citep{Blakeslee2001MNRAS.327.1004B,Jensen2001ApJ...550..503J,Tonry2001ApJ...546..681T} meaning that we can study the magnetic field with unprecedented spatial resolution.\\
In Section 2 we give a summary of the RM properties reported in detail in L25. In Section 3, we describe the methodology used to derive the intracluster magnetic field, while the corresponding results are presented in Section 4. Section 5 provides a discussion of the magnetic field properties of the Fornax cluster, placed in the context of a sample of 16 galaxy clusters and groups from the literature. The conclusions are provided in Section 6.

Throughout this paper we assume a $\Lambda$CDM cosmology with H$_0$ = 71 km\,s$^{-1}$Mpc$^{-1}$, $\Omega_m$=0.27, and $\Omega_{\Lambda}$= 0.73. At the distance of the Fornax cluster, 1\,arcsec corresponds to 0.1\,kpc.

\section{RM properties}
In L25 we constructed a dense ($\sim$80\,polarized sources/deg$^2$) RM grid by applying the RM-synthesis technique \citep{BrentjensdeBruyn} on the Stokes Q and U cubes with frequencies between 900\,MHz and 1.4\,GHz, 5\,MHz wide frequency channels and a spatial resolution of 13\,arcsec.
We identify the Faraday depth at the peak of the Faraday Dispersion Function as the observed RM.\\
A comparison between the RM obtained from the RM-synthesis and the one resulting from a $\lambda^2$-fit of the polarisation angle suggests that the majority of the sources are Faraday simple, i.e. the polarized signal of the sources is crossing a completely external Faraday screen. \\
At these frequencies the observed RM is primarily due to the Fornax cluster and the Galactic foreground.
Therefore we subtracted the Galactic foreground from the observed RM using the 2D-polynomial by \cite{anderson2021}.\\
The main RM features reported in detail in L25 are:
\begin{enumerate}
    \item a stripe of high-RM (in absolute value) sources crossing the cluster centre from N to SSW with positive and negative values going toward N and SSW respectively;
    \item a RM mean increment at a radial distance of about 300\,kpc from zero to $\sim$5\,rad m$^{-2}$;
    \item a $\sim$13\,rad m$^{-2}$ RM standard deviation plateau at large distances from the cluster centre.
\end{enumerate}
We suggested that the high-RM stripe is tracing the accretion of matter from the large scale structure surrounding the system: the Fornax cluster is linked with the Eridanus galaxy group on the North \citep{raj2024arXiv240703225R} and with the Fornax A group to the South-West \citep{Costa1988ApJ...327..544D}. The Eridanus group is at distance on the plane of the sky of around 14\,Mpc and at z=0.00557. Accretion events along the direction connecting the Fornax cluster and the Eridanus and Fornax A group can explain the enhancement of the RM values within the virial radius of the cluster.\\
The RM increase at about $r\sim$300\,kpc is likely due to the magnetic field power spectrum: in the eastern part sloshing motions are causing a decay of the magnetic field energy into small structures while in the western part the lack of dynamical motions is preserving a larger autocorrelation length; the combination of these effects results in a smaller RM in the eastern side compared to the western side of the cluster.\\
The RM standard deviation plateau at large distances from the cluster centre shows no significant directional dependence. We argue that this reflects a combination of direction-independent contributions that effectively set a noise floor:
the RM noise ($\sim$2\,rad m$^{-2}$), the local environment of the sources ($\sim$6\,rad m$^{-2}$), the Milky Way ($\sim$6\,rad m$^{-2}$), the cosmic web and the Fornax cluster (accounting for the remaining 9-10\,rad m$^{-2}$).

\section{Models and methods to measure the Fornax cluster's magnetic field}
In this Section we describe how we constrained the power spectrum, the strength and the profile of the magnetic field hosted by the Fornax cluster. \\
Due to the turbulent nature of the intra-cluster magnetic field it is important to consider different realisations of the same magnetic field model and to investigate which distribution has the higher probability to reproduce the data. The \texttt{FARADAY} tool \citep{murgia2004} was employed in this way in a number of works \citep[see e.g.][]{govoni2017} using Bayesian inference. \\
The basis of the Bayesian approach is to evaluate the posterior probability distribution P($\rm\theta$|D) of the model parameters $\rm\theta$ given the data D. The posterior probability distribution depends on the prior knowledge P($\rm\theta$) of model parameters, the likelihood P(D|$\rm\theta$) and the so-called evidence P(D):
\begin{equation}
  \rm  P(\theta | D) = \frac{P(D | \theta) P(\theta)}{P(D)},
\end{equation}
P(D) is defined as the integral of the likelihood function weighted by the prior over all the parameters' space:
\begin{equation}
  \rm  P(D) = \int P(D | \theta) P(\theta) d \theta.
\end{equation}
The posterior distribution is therefore an indicator of the most probable set of parameters.

We used a Monte-Carlo-Markov-Chain (MCMC) method with the Metropolis-Hastings algorithm to generate a sample of the posterior probability distribution without the need to calculate the evidence explicitly. The computation of the evidence is a hard task due to the fact that it is necessary to explore the entire parameter space.
The MCMC starts from a parameter set $\theta_0$ randomly extracted assuming a Gaussian prior distribution of all the parameters within a given range. 
The algorithm is run for several iterations by selecting new states according to a Q($\theta,\theta'$), a transitional kernel between the current and the proposed positions, i.e. $\theta$ and $\theta'$.
The proposed position is accepted with probability:
\begin{equation}
    h = min \left [ 1, \frac{ P(D|\theta')P(\theta')Q(\theta',\theta)}{P(D | \theta)P(\theta)Q(\theta,\theta')} \right ].
\end{equation}
As usual the MCMC starts with a number of "burn-in" steps during which the standard deviation of the transitional kernel, a multivariate Gaussian kernel in this work, is adjusted so that the average acceptance rate is 50\%. These steps are discarded from the accepted values and the posterior distribution is composed by the model parameters explored after the "burn-in" period. The final statistics on the model parameters is evaluated from the posterior distribution.

\subsection{Magnetic field power spectrum}
The magnetic field power spectrum is related to the RM power spectrum. In our analysis, we assumed a single power--law power spectrum for the magnetic field:
\begin{equation}
   \rm |B_k|^2 = A \cdot k^{-\zeta}
\end{equation}
with slope $\zeta$, normalisation $A$, fluctuating between a minimum and a maximum wave number $k_{\rm min}$ and $k_{\rm max}$ and zero elsewhere. In the physical space $k_{\rm min}$ and $k_{\rm max}$ are referred to as $\Lambda_{max}$ and $\Lambda_{min}$, with $\Lambda$=2$\pi/k$. \\
We used the Bayesian method to determine the best parameters for the assumed magnetic field model. Starting from a set of parameters $\zeta$, $A$, $\Lambda_{min}$ and $\Lambda_{max}$ the \texttt{FARADAY} tool can realize a simulated RM image, assuming proportionality between the 2D RM and the 3D magnetic field power spectra \citep{ensslin2003A&A...401..835E}. The RM noise associated with the data, computed following \cite{Sotomayor2013A&A...552A..58S}, was then added to this image. A weight equal to $\sqrt{2}$ for the cluster sources (which implies that these sources are in the cluster mid plane) was included. The simulated image was rescaled assuming a $\beta$-model distribution \citep{Cavaliere1976A&A....49..137C} for the thermal plasma density:
\begin{equation}
    n_e (r) = n_0 \left ( 1 + \frac{r^2}{r_c^2} \right )^{-3\beta/2}
    \label{eq:betamodel}
\end{equation}
with $n_0$ being the central plasma density, $r_c$ the cluster core radius. The most updated model we found in the literature for the Fornax cluster reports $\beta$=0.804, $r_c$=130\,kpc and $n_0$=1$\cdot 10^{-3}$ cm$^{-3}$ \citep{Reiprich2002ApJ...567..716R}, where $r_c$ and $n_0$ are corrected for the assumed cosmological model.
After blanking the simulated image in the same way as the observed RM image we computed the RM structure function $S(dr)$ and compared it with the observed RM structure function. This is a fundamental diagnostic to estimate the structure of the intra-cluster magnetic field. It is defined as the average mean square difference in RM values between pixels at a given distance $dr$:
\begin{equation}
    S(dr)=<[RM(r)-RM(r+dr)]^2>.
\end{equation}
The uncertainty associated with the RM structure function are computed by propagating the RM errors:
\begin{equation}
    \Delta S(dr)=\frac{2\sqrt{2S(dr)}}{\sqrt{N}} \Delta RM.
\end{equation}

\subsection{Magnetic field strength and trend}
Once the magnetic field power spectrum is constrained, we characterised the magnetic field strength using three dimensional numerical simulations. We assumed a radial profile for the magnetic field strength that scales with the thermal plasma distribution, following a power-law relation:
\begin{equation}
     B(r) = B_0 \left ( \frac{n_e(r)}{n_0} \right ) ^{\eta},
     \label{eq:br}
\end{equation}
where $B_0$ is the magnetic field strength at the cluster centre.
Each simulated cube was normalised assuming a specific value for $B_0$ and $\eta$. We then computed the RM by performing the integral in Eq. \ref{eq:rm}, assuming the $\beta$-model for the thermal plasma density reported previously.
Also in this case we took into account the RM noise, the weighting, and we blank the images as the observed RM image.
As already pointed out, at large distance from the cluster centre, the RM standard deviation tends to a value of $\sim$13\,rad m$^{-2}$ due to several contributions (noise, local environment, Milky Way, large scale structure, Fornax cluster). We added this extra-factor in the simulations.
The RM standard deviation radial profile was compared with the observed profile with the Bayesian approach described at the beginning of this Section.

\begin{figure*}
\centering
    \includegraphics[width=0.8\textwidth]{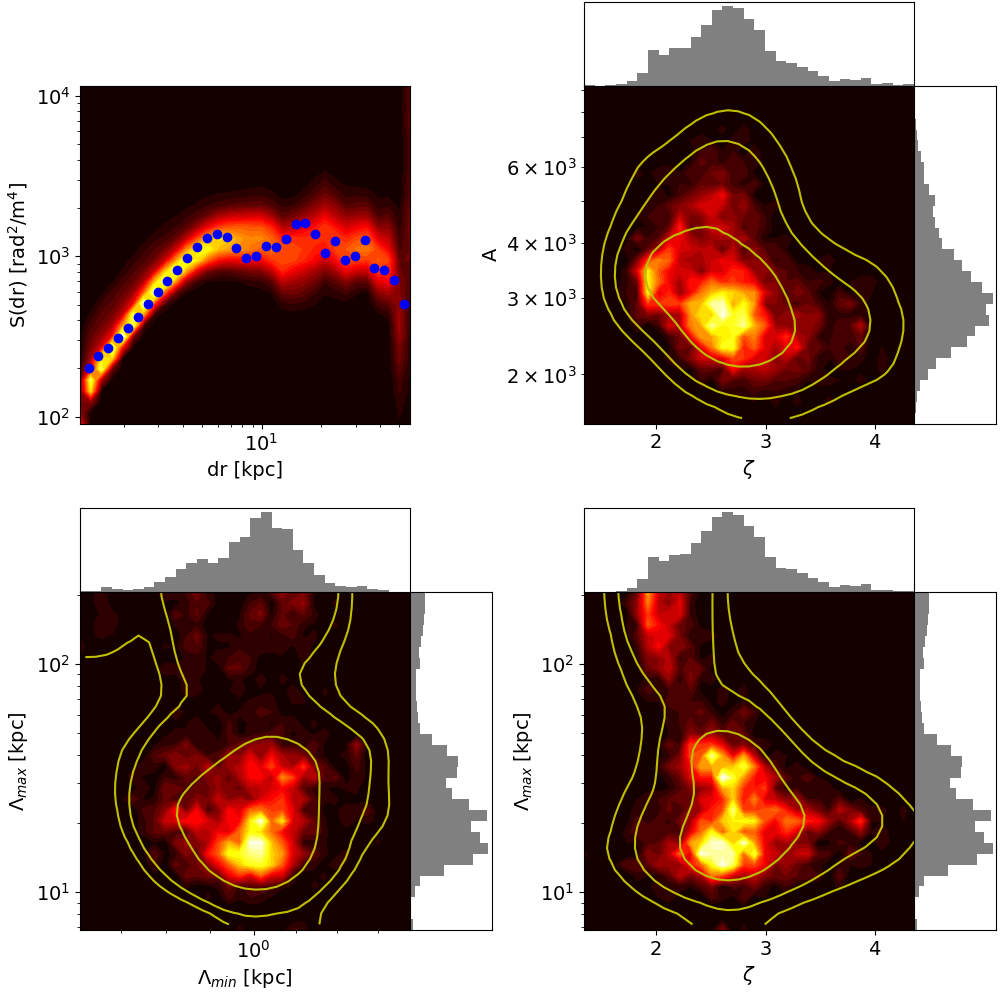}
    \caption{Results of the magnetic field power spectrum measurements. {\it Top left:} comparison between the observed RM structure function (blue points) and the distribution of the simulated RM structure functions. {\it Top right:} posterior distribution of the normalisation and slope of the magnetic field power spectrum. {\it Bottom left:} posterior distribution of the maximum and minimum scale of the magnetic field power spectrum. {\it Bottom right:} posterior distribution of the maximum scale and slope of the magnetic field power spectrum. In the last three panels we show the histogram of each parameter. The contours show the 1, 2 and 3$\sigma$ contours associated with the distributions.}
    \label{fig:2D}
\end{figure*}

\begin{figure*}
\centering
    \includegraphics[width=0.8\textwidth]{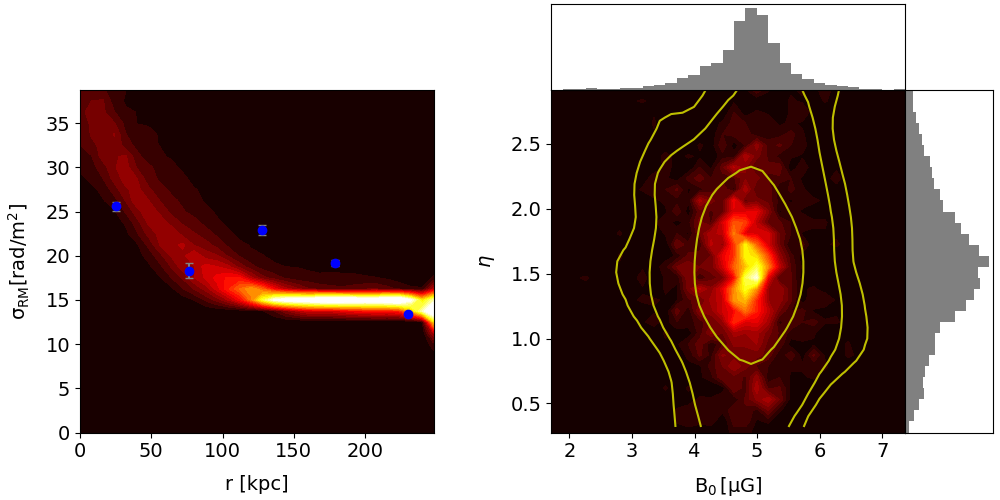}
    \caption{Results of the magnetic field profile measurements. {\it Left:} comparison between the observed RM standard deviation profile (blue points) and the distribution of the simulated ones. {\it Right:} posterior distribution of the central strength and slope of the magnetic field profile. }
    \label{fig:3D}
\end{figure*}

\section{Results}
To constrain the magnetic field properties we considered a field-of-view of 1.4\,degree ($\sim$512\,kpc) centred on the cluster centre. In this way the estimate should be less contaminated by the thermal plasma asymmetry \citep{su2017ApJ...851...69S, Reiprich2025arXiv250302884R}, for which we do not have an accurate model, and by the RM increment observed at a distance of about 300\,kpc from the cluster centre. 

\subsection{2D simulations}
We computed the RM structure function considering separations between pixels up to a maximum distance of 64\,kpc to be less sensitive to the density profile variations.    
We explored a parameter space for the slope $\zeta$ ranging from 0 to 5, for the minimum scale $\Lambda_{min}$ from 0.5 (i.e. two times the pixel size) to 2.2\,kpc and for the maximum scale $\Lambda_{max}$ from 2.5 to 230\,kpc. The scale limits are chosen according to the physical fluctuations that our data can probe, i.e. from half of the beam size to half of the field-of-view.
We also tested a broken power-law to model the magnetic field power spectrum. The results are consistent with what is reported below with the exception of the emerging of a strong degeneracy between the parameters. Indeed, a larger value for the maximum scale $\Lambda_{max}$ forces the power-law to be steeper to preserve the total power. The broken power-law collapses into a single power-law since at $\Lambda_{min}$ there is little contribution to the total power.\\
The results are shown in Fig. \ref{fig:2D}: the top right, bottom left and bottom right panels represent the posterior probability distributions of the normalisation vs slope, the maximum vs minimum scale and of the maximum scale vs slope respectively. On the top and on the right side of each panel an histogram shows the distribution of the corresponding parameter. The top left plot in Fig. \ref{fig:2D} displays the RM structure function models obtained with our simulations. The blue points are the measured RM structure function. After a steep and continuous increase the function starts oscillating from dr=6\,kpc around 1000 rad$^2$ m$^{-4}$. Beyond dr$\approx$30\,kpc the mean value of the structure function decreases to around 700 rad$^2$ m$^{-4}$.\\
At 68 per cent of confidence level, our modelling suggests a power-law power spectrum for the Fornax cluster's magnetic field having a slope of $\zeta$=2.7$^{+0.2}_{-0.4}$, a minimum physical scale $\Lambda_{min}$=(1.01$^{+0.01}_{-0.02}$)\,kpc, a maximum physical scale $\Lambda_{max}$=(15$^{+9}_{-2}$)\,kpc and a normalisation of 2990$^{+461}_{-683}$\,$\muup$G$^2$.
These scales correspond to a maximum and minimum scale of and $k_{max}$=6.2$\pm$1\,kpc$^{-1}$ and $k_{min}$=0.40$^{+0.02}_{-0.18}$\,kpc$^{-1}$ respectively. \\
We computed the magnetic field auto-correlation length using the best parameters for the magnetic field power-law power spectrum using the following equation \citep{ensslin2003A&A...401..835E}:
\begin{equation}
    \rm \Lambda_B = \frac{3\pi}{2} 
    \frac{\int_0^{\infty} |B_k|^2 k dk}{\int_0^{\infty} |B_k|^2 k^2 dk},
\end{equation}
and we find $\Lambda_B$=(3.3$^{+1.1}_{-0.3}$)\,kpc.\\

\subsection{3D simulations}
We determined the central strength and the power--law exponent of the Fornax cluster's magnetic field considering the RM standard deviation profile. The latter is computed on the image plane in 50\,kpc wide concentric anuli. 
We explored the parameter space between 0.01 and 50\,$\muup$G for B$_0$ and between -0.5 and 3 for $\rm \eta$. \\
The posterior probability distribution of B$_0$ and $\eta$ are shown on the right in Fig. \ref{fig:3D}. The left panel displays the comparison between the simulated and observed RM standard deviation radial profiles.\\
The best realisations of the magnetic field have a central strength of (5.0$^{+0.3}_{-0.4}$)\,$\muup$G decreasing with the thermal plasma density with a slope of $\eta$=(1.6$^{+0.3}_{-0.5}$). The observed RM standard deviation profile does not exhibit the monotonic radial decrease displayed by the simulated profiles. This behaviour has been observed in several cases, such as in 3C31, Hydra A \citep{laing2008}, Coma cluster \citep{bonafede2010}, 3C449 \citep{guidetti2010} A2345 \citep{stuardi2021}, and A194 \citep{govoni2017}.
In the latter case, the fluctuations are likely due to the smaller size of the central annulus compared to the outer annuli, leading to an under-sampling of the RM distribution; notably, simulations can reproduce the observed behaviour in the case of A194. In our study, the second radial bin is the least populated due to noise limitations (see Appendix B in L25), with 81, 15, 49, 96, and 82 independent measurements from the first to the last radial bin, respectively. Even with only 15 measurements, as in our most limited bin, we verified that the standard deviation of a Gaussian distribution can be correctly inferred through random extraction. Therefore, the number of sources in each bin is sufficient to represent the RM statistics, which explains why simulations cannot replicate the observed fluctuations.
To test if noise contributes to this effect, we applied a more stringent RM detection threshold, considering only measurements with an associated error smaller than 2 rad m$^{-2}$ (corresponding to polarized S/N > 13). The RM standard deviation values differ by about 1 rad m$^{-2}$ compared to those in Fig. \ref{fig:3D}, except in the second radial bin, where the value is 5 rad m$^{-2}$ lower. Even in this case, the number of independent measurements per bin remains significant (71, 7, 19, 46, 40 from the central to outer bins), although we observe a roughly 50\% depletion starting from the second bin outward.\\
There might be observational effects causing the observed scatter that we have not accounted for. However, numerical simulations \citep{Rappaz2024A&A...691A.132R,Loi2019MNRAS.490.4841L} show that the RM standard deviation radial profile can exhibit intrinsic scatter, particularly when the observations resolve the magnetic field fluctuation scale \citep[see Appendix A in][]{Rappaz2024A&A...691A.132R}. Such simulations incorporate more realistic models of the magnetic field structure and thermal plasma distribution, which could explain the variations in the RM structure function.\\
In any case, considering that our simulations rely on simple models for the RM that, for instance cannot reproduce the observed thermal plasma asymmetry, we find the results presented in Fig. \ref{fig:3D} to be quite satisfactory.

\section{Comparison with the literature and discussion}
This paper presents the first accurate measurement of the Fornax cluster's magnetic field. We noted that the magnetic field strength derived in this work is consistent with the 10\,$\muup$G upper limit established by \cite{su2017ApJ...851...69S}, which was based on the modelling of a candidate Kelvin--Helmholtz instability eddy detected 3\,kpc from the cluster centre.
To better understand the implications of our findings for overall cluster physics, we perform a qualitative comparison with the properties of magnetic fields measured in galaxy clusters and groups.\\
Table \ref{tab:literature} summarizes the magnetic field estimates available in the literature, derived from spectro-polarimetric modelling of cluster diffuse sources (e.g., A523, A665) or RM analysis (e.g., the remaining systems) using various numerical tools and assumptions. Magnetic field parameters that were fixed a priori in the original modelling are marked with an asterisk. All values have been rescaled to the cosmological model adopted in this study. This collection encompasses galaxy clusters and groups in diverse dynamical states.
Massive relaxed clusters are the end products of low-mass systems that have accreted mass through numerous merging events. These events are thought to drive the amplification of intra-cluster magnetic fields on Gyrs timescales \citep[see e.g.][]{Subramanian2006MNRAS.366.1437S}, a duration that corresponds to the typical relaxation period of galaxy clusters \citep{Zhang2022MNRAS.516...26Z}. 
For the purposes of our analysis, it is crucial to distinguish between merging and relaxed clusters. In the latter, we expect higher magnetic field strengths and smaller auto-correlation length, as magnetic energy cascades from large injection scale down to smaller scales. Conversely, minor mergers and the radio activity promoted by cluster AGN can induce instabilities that amplify magnetic fields more rapidly \citep{paola2019MNRAS.486..623D}. The presence of extended central radio galaxies is highlighted throughout our discussion of the results.
At lower masses ($\sim$10$^{13}$M$_{\odot}$), the magnetisation process remains poorly understood. Such systems are characterized by smaller sizes and lower thermal densities, which implies: (a) potential departures from the magneto-hydro-dynamical (MHD) approach used in cosmological numerical simulations due to the collisionless nature of the thermal plasma \citep{Simionescu2019SSRv..215...24S}, and (b) a more prominent role of the AGN feedback \citep[see e.g.][]{Eckert2021Univ....7..142E}. To investigate if and how magnetic field properties reflect the dynamical state of the systems, their physics and the role of central powerful radio sources, we adopt the following categories: 
\begin{enumerate}
    \item Massive Merging Clusters (MMC) and Massive Relaxed Clusters (MRC) for systems with masses within R500\footnote{R$_{500}$ is the radius of a spherical region within a galaxy cluster that has a mean density of 500 times the critical density of the universe at the cluster's redshift.} M$_{500}$>10$^{14}$M$_{\odot}$;
    \item Low-mass Merging Clusters (LMC) and Low-mass Relaxed Clusters (LRC) for systems with M$_{500}$<10$^{14}$M$_{\odot}$;
    \item Galaxy Groups (GG) based on existing literature classification. 
\end{enumerate}
We adopted the M$_{500}$ and R$_{500}$ values from the Meta Catalogue X--ray Galaxy Clusters \citep[MCXC,][]{piffaretti2011}. The Fornax Cluster presents an intermediate dynamical state, since it is a moderate cool-core cluster with a merging going on with the Fornax A group. Our analysis is limited to the central part of the cluster, where X-ray observations show no evidence of merger-induced shock waves. Consequently, we have classified it as an LRC. In the following sections, we discuss the main properties of the sample reported in Table \ref{tab:literature}, which comprises 17 galaxy clusters and groups. \\

\begin{figure}
    \centering
    \includegraphics[width=0.47\textwidth]{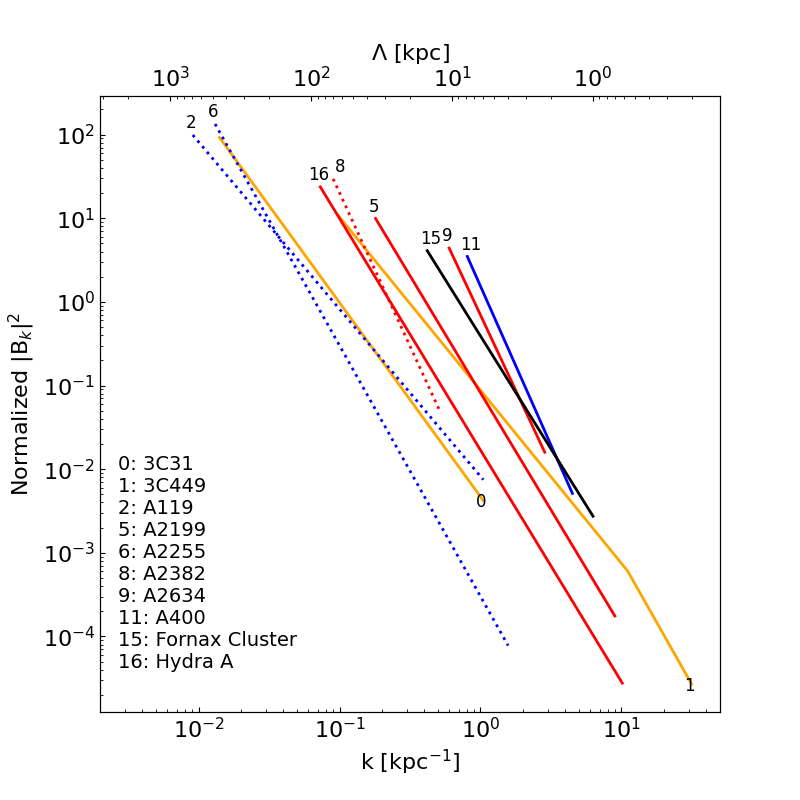}
    \caption{Normalized magnetic field power spectra: blue, red, and orange corresponds to merging galaxy clusters, relaxed galaxy clusters, and galaxy groups, respectively; solid and dotted lines are used for systems with or without central extended radio galaxies. The Fornax cluster is represented by a black solid line.}
    \label{fig:ps}
\end{figure}
\begin{figure}
    \centering
    \includegraphics[width=0.47\textwidth]{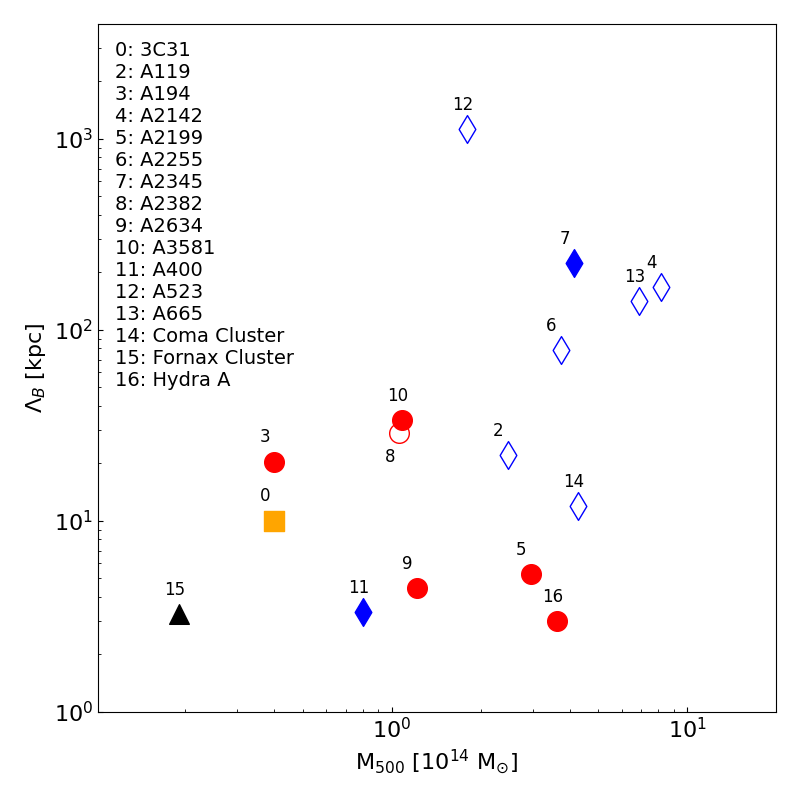}
    \caption{Magnetic field auto-correlation length as a function of M$_{500}$ colour coded as Fig. \ref{fig:ps}. Diamonds, circles and squares show merging, relaxed clusters and galaxy groups, respectively. The Fornax cluster measurement is shown as a black triangle. Filled symbols indicate which systems host extended radio galaxy at the centre.}
    \label{fig:lb}
\end{figure}

\subsection{The shape of the magnetic field power spectrum}
\label{sect:bpower}
Fig. \ref{fig:ps} illustrates the magnetic field power spectra for our sample, plotted using the minimum and maximum fluctuation scales inferred in each respective study. We include only those measurements obtained without 'a priori' assumptions regarding the power spectrum slope.\\
A Kolmogorov-like spectrum is observed in A2382. Consistent values are found by \cite{vogt2003A&A...412..373V} for A2634 and A400. In contrast, we find $\zeta\approx$2 for A119, 3C449, $\zeta\approx$2.7 for 3C31, $\zeta\approx$2.7 for Hydra A, the Fornax cluster, and A2199, $\zeta\approx$3 for A2255. 
MHD simulations incorporating AGN feedback \citep[see][]{Tevlin2025A&A...701A.114T}, typically show a Kolmogorov spectrum at small scales ($\leq100$\,kpc), while at large scales ($\geq100$\,kpc) a k$^{-1/2}$ Kazantsev trend emerges due to dynamo activity. Consequently, the observed deviations from a Kolmogorov spectrum may represent a compromise between these two regimes—leading to a flatter overall spectrum—or they may highlight the resolution limits of current MHD simulations. Indeed, accurately reconstructing galaxy cluster physics from the smallest scales (e.g., supernova-driven turbulence) to the largest scales (above a few Mpc) remains computationally challenging. The former scenario might apply to A119 and A2255, although in the latter case, the power spectrum was derived fixing the maximum scale of fluctuation to 512\,kpc; adopting smaller or larger values would result in a flatter or steeper spectrum, respectively.
In most other cases, the large scales are not sufficiently sampled. As discussed in Section 4 for the Fornax cluster, a broken power-law model, which in principle could probe both trends, effectively collapsed into a single power-law.
Alternatively, the magnetic field in these systems may not yet have reached the saturation phase \citep{Schekochihin2004ApJ...612..276S}. This would challenge recent theoretical results suggesting that saturation is achieved as early as \citep[z$\approx$4,][]{Tevlin2025A&A...701A.114T}. 
We acknowledge the strong degeneracy between power spectrum parameters, the inherent simplicity of our assumed models, and possible bias due to the comparison of results derived using different approaches. Nevertheless, these results underscore the complex physics governing these environments, which can shape the magnetic field power spectrum and cause a flattening relative to the Kolmogorov expectation. This suggests that more power is injected at small scales than currently predicted by state-of-the-art simulations and/or that the magnetic field is still undergoing active amplification.\\
It is instructive to examine the magnetic field auto-correlation lengths $\Lambda_B$, which are expected to reflect the characteristic turbulence scales of the ICM. In Fig. \ref{fig:lb}, we plot this quantity as a function of cluster mass. Massive merging clusters consistently exhibit larger auto-correlation lengths compared to massive relaxed systems, regardless of the overall cluster mass. This difference likely arises because, in relaxed clusters, the magnetic field energy has had sufficient time to cascade from large injection scales down to smaller scales, as previously noted by Vacca et al. (2026). 
Interestingly, low-mass galaxy clusters and groups display auto-correlation lengths similar to those of massive relaxed clusters. Even lower values are observed in the Fornax cluster and the merging low-mass cluster A400. Notably, both of these systems host an extended central radio galaxy, which could be responsible for more rapid and localized magnetic field amplification at small scales. This mechanism will be discussed in further detail in the following Section.
\begin{figure}
    \centering
    \includegraphics[width=0.47\textwidth]{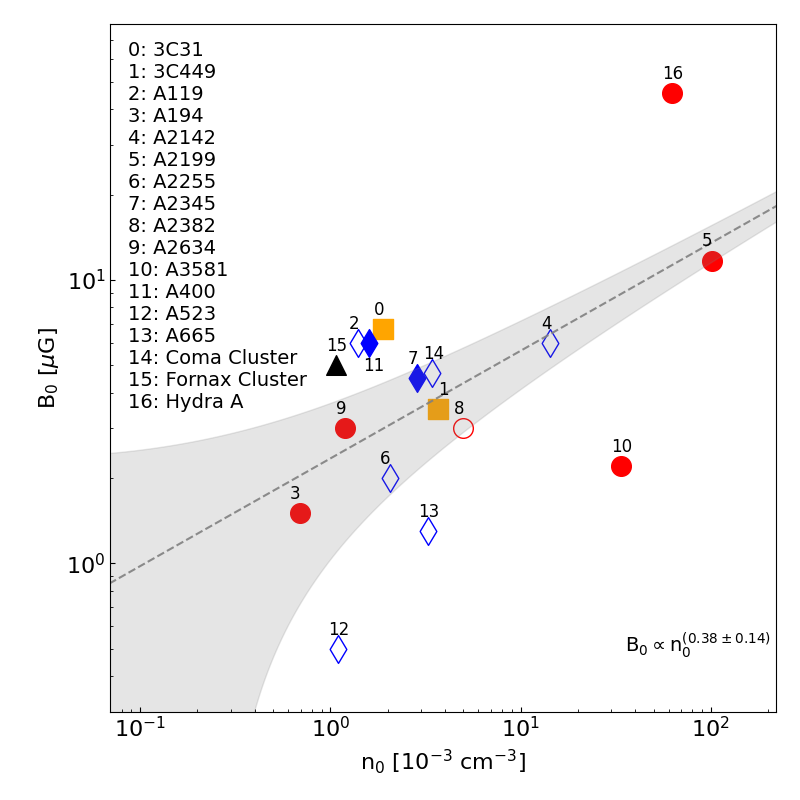}
    \caption{Magnetic field central strength as a function of the central thermal density with same colour and symbol code described in Fig. \ref{fig:lb}. The dashed line represents the best-fit model (parameters in the bottom right corner), with uncertainty indicated by the gray shaded area.}
    \label{fig:B_corr}
\end{figure}
\begin{figure}
    \centering
    \includegraphics[width=0.47\textwidth]{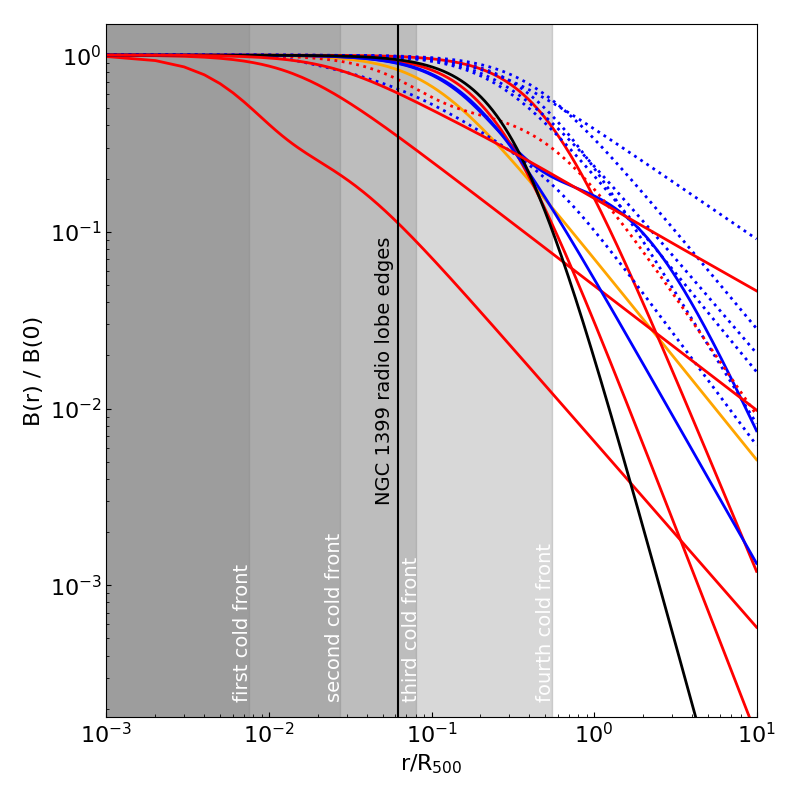}
    \caption{Normalized magnetic field strength profiles as a function of the ratio between the distance from the cluster centre and R$_{500}$. We used the same line styles as explained in Fig. \ref{fig:ps}. The gray shaded areas indicate the location of the cold front detected by \citet{su2017ApJ...851...69S} while a vertical black line shows the extension of the radio lobes associated with NGC\,1399.}
    \label{fig:B_profs}
\end{figure}
\subsection{Magnetic field values and profiles}
Fig. \ref{fig:B_corr} displays the magnetic field central strength as a function of the central plasma density. Using a linear fit in the log--log plane, we find a scaling relation of $B_0 \propto n_0^{(0.38\pm0.14)}$. The fit uncertainties, derived by propagating the errors on the fit parameters, are indicated by the shaded gray area. This plot represents an update to the results reported by \cite{govoni2017} and Vacca et al. (2026); unlike those studies, we include measurements of A400, A2634, A2345, A2142, A3581, and the Fornax cluster, while the latter work also incorporates A665 and A523 measurements. \cite{govoni2017} and Vacca et al. (2026) found $B_0 \propto n_0^{0.47}$ and $B_0 = n_0^{(0.56\pm0.17)}$, respectively. Our flatter correlation remains consistent with these previous works within the uncertainties.\\
Notably, our target exhibits a central magnetic field strength approximately twice as high as predicted by the fitted correlation. 
Similar excesses are observed in Hydra A, 3C31, A400, and A119, whereas A665, A523, and A3581 are characterized by lower values. A key distinction among these systems is the presence of one or more extended radio sources at the centre of the first group (with the exception of A119).\\
To further investigate this, Fig. \ref{fig:B_profs} shows the magnetic field radial profile as a function of the ratio between the distance from the cluster centre and R500.
We find that the Fornax cluster exhibits the steepest radial profile among all systems considered in this study, suggesting that an efficient magnetic field amplification mechanism operates near to the cluster centre. MHD simulations demonstrate that cold fronts can significantly enhance magnetic fields along the boundaries of sloshing regions, thereby steepening the radial profile \citep[e.g.][]{ZuHone2011ApJ...743...16Z}. Indeed, a series of cold fronts induced by the infall of NGC\,1404 has been detected in Fornax at radii of approximately 3, 11, 32, and 220\,kpc \citep[][converted in our cosmology]{su2017ApJ...851...69S}. These are indicated in Fig. \ref{fig:B_profs} by vertical gray shaded areas. However, while the magnetic field strength remains high in the inner regions, it drops to 10\% on its central value at 220\,kpc. This decline suggests that additional mechanisms may contribute to the amplification of the intracluster magnetic field.\\ 
The central AGN activity, which has produced two lobes extending up to $\sim$25\,kpc, could be a driver of recent re--amplification.
In low-density environments such as the ICM of Fornax, AGN outbursts can efficiently amplify magnetic fields through shocks generated by the inflation of radio bubbles \citep[see e.g.][and reference therein]{Donnert2018SSRv..214..122D}. The observed steep profile and the high central strength can thus be explained by the interaction between the extended central radio source and the surrounding ICM. This interpretation is supported by the chronology of events: AGN activity is more recent than the sloshing episode. The symmetry of the two radio lobes implies that their inflation occurred late in the cluster's evolution; otherwise, sloshing would have significantly distorted the radio morphology \citep{paola2024ApJ...977..221D}. Moreover, \citet{su2017ApJ...851...69S} estimate the sloshing age to be of order Gyr, far exceeding the typical radiative lifetime of AGN. \\
Simulations by \citet{paola2024ApJ...977..221D}, combining binary cluster mergers with bidirectional central jet injection, strengthen this scenario. At early stages (see Fig. 5 in that paper), when the AGN is active but before the merger has disrupted the system, the central magnetic field strength exceeds the overall trend observed once sloshing has mixed the radio lobes with the ICM. These results point to AGN-driven amplification as the origin of the steep profile in Fornax. This factor, combined with pre-amplification by sloshing motions, likely explains the cluster's relatively high central magnetic field strength.\\
Prior to this work, the steepest reported power--law exponent was $\eta \approx$ 1.1, measured in the low-mass cluster A194. Similar to Fornax, A194 hosts a central AGN with symmetric jets and has a relatively low central density; however, unlike our target, it lacks evidence of cluster mergers or cold fronts. In that case, AGN outbursts likely dominate the central magnetic field amplification.
Furthermore, the most prominent departures from the correlation in Fig. \ref{fig:B_corr} are consistent with our proposed interpretation: higher central strength are found in systems hosting one or more extended radio sources. A119 is a notable exception, but as reported by \cite{Watson2023ApJ...955..103W}, it is characterized by recent and on-going merger activity, which could justify the observed magnetic field amplification across the cluster volume, resulting in a flat radio profile. \\
Among the systems falling below the B$_0$--n$_0$ correlation, A3581 stands out as an unexpected outlier. Given its relaxed dynamical state, one would expect a central magnetic field strength on the order of tens of $\muup$G. The lower value reported by \cite{khadir2025arXiv251118532K} may result from the decision to constrain the power-law exponent parameter ($\eta$) to values below 0.5. While this was motivated by the observed flattening of the RM profile, Table \ref{tab:literature} shows that other relaxed systems are well-described by profiles with $\eta = 1$. Adopting this higher value would increase the central strength to $\approx$37$\muup$G, bringing A3581 into alignment with the other upper outliers.\\
A similar discrepancy may affect A665 \citep{vacca2010A&A...514A..71V}, for which $\eta$ was derived from the radio halo surface brightness radial profile under the assumption of equipartition between the magnetic field and relativistic particles. This assumption implies a steep scaling between radio emissivity $J_{\nu}$ and the magnetic field ($J_{\nu} \propto B^4$), leading to the profile $J_{\nu} \propto (1 + (r/r_c)^2)^{-6\beta\eta}$ (assuming Eq. \ref{eq:br}). However, since the equipartition condition is physically difficult to achieve, a scaling of $J_{\nu} \propto B^2$ may be more plausible and $\eta$ would be $\approx 1$, yielding $B_0 \approx 3$ $\muup$G, consistent with the established $B_0$–$n_0$ correlation. The final outlier in the low-$B_0$ regime is A523. The model reported by \citep{Vacca2022MNRAS.514.4969V} was specifically designed to demonstrate that an intracluster magnetic field can generate polarized radio halo emission on scales exceeding those of the associated total intensity. Because the primary objective was to reproduce these polarized structures, the ($B_0, \eta$) parameter space was not explored in detail. Consequently, this measurement likely suffers from the same modelling biases discussed above, where a different choice of $\eta$ could bring the cluster into better agreement with the $B_0$–$n_0$ correlation. It is important to note that the correlation presented by Vacca et al. (2026) remains statistically consistent with the measurements for A665 and A523 within their reported uncertainties. While models utilizing a curved power spectrum could introduce systematic offsets, our primary conclusions remain robust even if A2142 and A2345 are excluded.\\
Our results establish a crucial benchmark for cluster magnetisation, yet they also reveal that a consistent and homogeneous modelling strategy is required across the literature to ensure statistically robust and bias-free comparisons between different environments.

A more quantitative study of magnetic field properties in galaxy clusters was recently conducted by 
\citet{Osinga2025A&A...694A..44O}. Through a statistical analysis of 124 massive Planck clusters at low redshift (z < 0.35) using depolarisation and RM stacking techniques, they found that the best--fit RM standard deviation profiles are obtained with $\eta$ < 0.5
However, they also noted that more advanced modelling of the intracluster magnetic field is required to further refine this comparison. Furthermore, they reported flatter RM profiles in cool-core clusters compared to non-cool-core systems. Since Fornax is a moderate cool-core cluster, it would be expected to exhibit a relatively flat profile. The fact that this is not observed suggests that magnetic field amplification and distribution in clusters depend strongly on their dynamical state, mass, and density, as well as the presence of powerful AGN.

In addition, we remark that the time scales of merging events and AGN outburst cycles differ significantly, spanning Gyrs versus tens or hundreds Myrs, respectively. While widespread magnetic field amplification occurs slowly via turbulence induced by merging events, more rapid and localized amplification is expected when a central AGN is radio active \citep{paola2019MNRAS.486..623D}. Indeed, flat profiles ($\eta$<0.5) are found in the systems forming the Shapley Supercluster Core \citep{alonso2026A&A...705A.143A}, which consists of two massive galaxy clusters connected by an intracluster bridge detected with Planck \citep{planck2016A&A...594A..22P} and incorporates two massive groups of galaxies. This system is characterized by large scale diffuse radio emission \citep{venturi2022A&A...660A..81V}, a clear tracer of an on-going merger event. 
Finally, we note that the correlation found by \cite{balboni2025A&A...695A.180B} between the magnetic field and the galaxy cluster mass (B$\propto$M$^b$, with b$\approx$2) describes only A2255 and the Coma Cluster within our sample, at least within the uncertainties. This is perhaps unsurprising, as the correlation was derived from a sample of 18 massive galaxy clusters (M$_{500}\geq$2.6$\cdot$10$^{14}$M$_{\odot}$) hosting diffuse radio sources. This discrepancy further highlights the need for deep investigations of the complex physics at play in the different environments that characterized both galaxy groups and clusters.

\section{Conclusions}
We present the first detailed measurements of the Fornax cluster's magnetic field, up to a radius of about 260\,kpc ($\sim$0.37$\times$R$_{vir}$), alongside a qualitative discussion on magnetisation across a sample of 17 galaxy clusters and groups.
Exploiting the dense RM grid provided by the MeerKAT Fornax Survey, we reconstructed the properties of the intracluster magnetic field by comparing observed RM data with numerical simulations using a Bayesian approach. 
To avoid potential contamination from the evident asymmetry in the X-ray distribution, for which an accurate reconstruction is currently unavailable, we restricted the RM measurements to within 250 kpc of the cluster centre.
We modelled the power spectrum with a power-law, finding a slope of 2.7$^{+0.2}_{-0.4}$ and a minimum and a maximum scale of 1.01$^{+0.01}_{-0.02}$ and 15$^{+9}_{-2}$\,kpc, respectively. The resulting magnetic field auto-correlation length was determined to be $\Lambda_B$=3.3$^{+1.1}_{-0.3}$\,kpc. 
By assuming that the magnetic field scales radially with the thermal plasma density, we constrained its central strength to 5.0$^{+0.3}_{-0.4}$\,$\muup$G and derived a steep a power--law exponent of $\eta$=1.6$^{+0.3}_{-0.5}$. Compared to other clusters in the literature, Fornax exhibits a remarkably steep magnetic field radial profile and a relatively high central strength.\\
Our qualitative analysis of 17 systems studied via RM in the literature reveals that, in agreement with previous work, massive merging clusters generally host larger auto-correlation lengths (Fig. \ref{fig:lb}). Conversely, massive relaxed clusters and low-mass clusters/galaxy groups are characterized by smaller auto-correlation lengths.
Notably, among the 10 systems with an inferred power spectrum slope, only three exhibit a Kolmogorov spectrum, while the remainder show a flatter behaviour (Fig. \ref{fig:ps}). These findings suggest that state-of-the art numerical simulations may still lack magnetic field power at kpc-scales or that the magnetic field is undergoing ongoing amplification.
At the cluster centre, the sample follows the relation B$_0\propto$ n$_0^{(0.38\pm0.14)}$ (Fig. \ref{fig:B_corr}), consistent within the uncertainties with previous studies \citep[e.g.][Vacca et al. 2026]{govoni2017}; however, several systems, including Fornax, display magnetic field strengths approximately twice as high as predicted by this correlation. \\
Since Fornax is a low mass environment hosting a radio-loud central BCG, we conclude that the amplification of its intra-cluster magnetic field likely stems from a combination of sloshing-induced turbulence and, more recently, radio galaxy activity. This interpretation is supported by the steep radial slope of the magnetic field profile (Fig. \ref{fig:B_profs}). Furthermore, four of five upper outliers in the B$_0$--n$_0$ correlation, namely Hydra A, A400, 3C31, and the Fornax cluster, host one or more extended radio galaxies at their centres. 
Their interaction with the environment may have enhanced the magnetic field amplification at small scales, as evidenced by the small auto-correlation lengths observed in these systems (see Fig. \ref{fig:lb}).\\
This study enhances our understanding of cosmic magnetism and sheds light on the amplification mechanisms operating across diverse large-scale environments. Future systematic surveys with SKA precursors and pathfinders, and eventually with the SKA-Mid telescope (Loi et al. 2026), will be essential to clarify how dynamical state, mass, density, and AGN feedback shape the observed magnetic fields and influence the evolution of cluster-embedded galaxies.

\begin{acknowledgements}
The authors thank the anonymous referee for their helpful comments that improved the manuscript.
We are grateful to the full MeerKAT team for their work building, commissioning and operating MeerKAT, and for their support to the MeerKAT Fornax Survey.
The MeerKAT telescope is operated by the South African Radio Astronomy Observatory, which is a facility of the National Research Foundation, an agency of the Department of Science and Innovation.
We acknowledge  support from MAEC grant nr ZA23GR03 (RADIOMAP).
This project has received funding from the European Research Council (ERC) under the European Union’s Horizon 2020 research and innovation programme (grant agreement no. 679627, “FORNAX”; and grant agreement no. 882793, “MeerGas”). 
The data of the MeerKAT Fornax Survey are reduced using the CARACal pipeline, partially supported by ERC Starting grant number 679627, MAECI Grant Number ZA18GR02, DST-NRF Grant Number 113121 as part of the ISARP Joint Research Scheme, and BMBF project 05A17PC2 for D-MeerKAT. Information about CARACal can be obtained online under the URL: https://caracal.readthedocs.io. 
The INAF - OAC computer cluster used in this work has been acquired within a project aimed to enhance the Sardinia Radio Telescope (SRT). The Enhancement of the SRT for the study of the Universe at high radio frequencies is financially supported by the National Operative Program (Programma Operativo Nazionale - PON) of the Italian Ministry of University and Research "Research and Innovation 2014-2020", Notice D.D. 424 of 28/02/2018 for the granting of funding aimed at strengthening research infrastructures, in implementation of the Action II.1 – Project Proposal PIR01$\_$00010.
This work was carried out thanks to the funding of the Regione Autonoma della Sardegna, ai sensi della Legge Regionale 7 agosto 2007, n.7 "Promozione della Ricerca Scientifica e dell'Innovazione Tecnologica in Sardegna".
FL acknowledges financial support from the Italian Ministry of University and Research – Project Proposal CIR01-00010.
VV acknowledges support from the Premio per Giovani Ricercatori "Gianni Tofani" II edizione, promoted by INAF-Osservatorio Astrofisico di Arcetri (DD n. 84/2023).
PK is partially supported by the BMBF project 05A23PC1 for D-MeerKAT III.
FMM carried out part of the research activities described in this paper with contribution of the Next Generation EU funds within the National Recovery and Resilience Plan (PNRR), Mission 4 - Education and Research, Component 2 - From Research to Business (M4C2), Investment Line 3.1 - Strengthening and creation of Research Infrastructures, Project IR0000034 – "STILES - Strengthening the Italian Leadership in ELT and SKA".
This work is based on data from eROSITA, the soft X-ray instrument aboard SRG, a joint Russian-German science mission supported by the Russian Space Agency (Roskosmos), in the interests of the Russian Academy of Sciences represented by its Space Research Institute (IKI), and the Deutsches Zentrum für Luft- und Raumfahrt (DLR). The SRG spacecraft was built by Lavochkin Association (NPOL) and its subcontractors, and is operated by NPOL with support from the Max Planck Institute for Extraterrestrial Physics (MPE). The development and construction of the eROSITA X-ray instrument was led by MPE, with contributions from the Dr. Karl Remeis Observatory Bamberg \& ECAP (FAU Erlangen-Nuernberg), the University of Hamburg Observatory, the Leibniz Institute for Astrophysics Potsdam (AIP), and the Institute for Astronomy and Astrophysics of the University of Tübingen, with the support of DLR and the Max Planck Society. The Argelander Institute for Astronomy of the University of Bonn and the Ludwig Maximilians Universität Munich also participated in the science preparation for eROSITA.
The eROSITA data shown here were processed using the eSASS software system developed by the German eROSITA consortium.
\end{acknowledgements}
\bibliographystyle{aa}
\bibliography{fornax}

@ARTICLE{planck2016A&A...594A..22P,
       author = {{Planck Collaboration} and {Aghanim}, N. and {Arnaud}, M. and {Ashdown}, M. and {Aumont}, J. and {Baccigalupi}, C. and {Banday}, A.~J. and {Barreiro}, R.~B. and {Bartlett}, J.~G. and {Bartolo}, N. and {Battaner}, E. and {Battye}, R. and {Benabed}, K. and {Beno{\^\i}t}, A. and {Benoit-L{\'e}vy}, A. and {Bernard}, J.-P. and {Bersanelli}, M. and {Bielewicz}, P. and {Bock}, J.~J. and {Bonaldi}, A. and {Bonavera}, L. and {Bond}, J.~R. and {Borrill}, J. and {Bouchet}, F.~R. and {Burigana}, C. and {Butler}, R.~C. and {Calabrese}, E. and {Cardoso}, J.-F. and {Catalano}, A. and {Challinor}, A. and {Chiang}, H.~C. and {Christensen}, P.~R. and {Churazov}, E. and {Clements}, D.~L. and {Colombo}, L.~P.~L. and {Combet}, C. and {Comis}, B. and {Coulais}, A. and {Crill}, B.~P. and {Curto}, A. and {Cuttaia}, F. and {Danese}, L. and {Davies}, R.~D. and {Davis}, R.~J. and {de Bernardis}, P. and {de Rosa}, A. and {de Zotti}, G. and {Delabrouille}, J. and {D{\'e}sert}, F.-X. and {Dickinson}, C. and {Diego}, J.~M. and {Dolag}, K. and {Dole}, H. and {Donzelli}, S. and {Dor{\'e}}, O. and {Douspis}, M. and {Ducout}, A. and {Dupac}, X. and {Efstathiou}, G. and {Elsner}, F. and {En{\ss}lin}, T.~A. and {Eriksen}, H.~K. and {Fergusson}, J. and {Finelli}, F. and {Forni}, O. and {Frailis}, M. and {Fraisse}, A.~A. and {Franceschi}, E. and {Frejsel}, A. and {Galeotta}, S. and {Galli}, S. and {Ganga}, K. and {G{\'e}nova-Santos}, R.~T. and {Giard}, M. and {Gonz{\'a}lez-Nuevo}, J. and {G{\'o}rski}, K.~M. and {Gregorio}, A. and {Gruppuso}, A. and {Gudmundsson}, J.~E. and {Hansen}, F.~K. and {Harrison}, D.~L. and {Henrot-Versill{\'e}}, S. and {Hern{\'a}ndez-Monteagudo}, C. and {Herranz}, D. and {Hildebrandt}, S.~R. and {Hivon}, E. and {Holmes}, W.~A. and {Hornstrup}, A. and {Huffenberger}, K.~M. and {Hurier}, G. and {Jaffe}, A.~H. and {Jones}, W.~C. and {Juvela}, M. and {Keih{\"a}nen}, E. and {Keskitalo}, R. and {Kneissl}, R. and {Knoche}, J. and {Kunz}, M. and {Kurki-Suonio}, H. and {Lacasa}, F. and {Lagache}, G. and {L{\"a}hteenm{\"a}ki}, A. and {Lamarre}, J.-M. and {Lasenby}, A. and {Lattanzi}, M. and {Leonardi}, R. and {Lesgourgues}, J. and {Levrier}, F. and {Liguori}, M. and {Lilje}, P.~B. and {Linden-V{\o}rnle}, M. and {L{\'o}pez-Caniego}, M. and {Mac{\'\i}as-P{\'e}rez}, J.~F. and {Maffei}, B. and {Maggio}, G. and {Maino}, D. and {Mandolesi}, N. and {Mangilli}, A. and {Maris}, M. and {Martin}, P.~G. and {Mart{\'\i}nez-Gonz{\'a}lez}, E. and {Masi}, S. and {Matarrese}, S. and {Melchiorri}, A. and {Melin}, J.-B. and {Migliaccio}, M. and {Miville-Desch{\^e}nes}, M.-A. and {Moneti}, A. and {Montier}, L. and {Morgante}, G. and {Mortlock}, D. and {Munshi}, D. and {Murphy}, J.~A. and {Naselsky}, P. and {Nati}, F. and {Natoli}, P. and {Noviello}, F. and {Novikov}, D. and {Novikov}, I. and {Paci}, F. and {Pagano}, L. and {Pajot}, F. and {Paoletti}, D. and {Pasian}, F. and {Patanchon}, G. and {Perdereau}, O. and {Perotto}, L. and {Pettorino}, V. and {Piacentini}, F. and {Piat}, M. and {Pierpaoli}, E. and {Pietrobon}, D. and {Plaszczynski}, S. and {Pointecouteau}, E. and {Polenta}, G. and {Ponthieu}, N. and {Pratt}, G.~W. and {Prunet}, S. and {Puget}, J.-L. and {Rachen}, J.~P. and {Reinecke}, M. and {Remazeilles}, M. and {Renault}, C. and {Renzi}, A. and {Ristorcelli}, I. and {Rocha}, G. and {Rossetti}, M. and {Roudier}, G. and {Rubi{\~n}o-Mart{\'\i}n}, J.~A. and {Rusholme}, B. and {Sandri}, M. and {Santos}, D. and {Sauv{\'e}}, A. and {Savelainen}, M. and {Savini}, G. and {Scott}, D. and {Spencer}, L.~D. and {Stolyarov}, V. and {Stompor}, R. and {Sunyaev}, R. and {Sutton}, D. and {Suur-Uski}, A.-S. and {Sygnet}, J.-F. and {Tauber}, J.~A. and {Terenzi}, L. and {Toffolatti}, L. and {Tomasi}, M. and {Tramonte}, D. and {Tristram}, M. and {Tucci}, M. and {Tuovinen}, J. and {Valenziano}, L. and {Valiviita}, J. and {Van Tent}, B. and {Vielva}, P. and {Villa}, F. and {Wade}, L.~A. and {Wandelt}, B.~D. and {Wehus}, I.~K. and {Yvon}, D.},
        title = "{Planck 2015 results. XXII. A map of the thermal Sunyaev-Zeldovich effect}",
      journal = {\aap},
         year = 2016,
        month = sep,
       volume = {594},
          eid = {A22},
        pages = {A22},
          doi = {10.1051/0004-6361/201525826},
archivePrefix = {arXiv},
       eprint = {1502.01596},
 primaryClass = {astro-ph.CO},
       adsurl = {https://ui.adsabs.harvard.edu/abs/2016A&A...594A..22P}
}

@ARTICLE{alonso2026A&A...705A.143A,
       author = {{Alonso-L{\'o}pez}, D. and {O'Sullivan}, S.~P. and {Bonafede}, A. and {B{\"o}ss}, L.~M. and {Stuardi}, C. and {Osinga}, E. and {Anderson}, C.~S. and {Van Eck}, C.~L. and {Carretti}, E. and {West}, J.~L. and {Akahori}, T. and {Dolag}, K. and {Giacintucci}, S. and {Khadir}, A. and {Ma}, Y.~K. and {Malik}, S. and {McClure-Griffiths}, N. and {Rudnick}, L. and {Seidel}, B.~A. and {Tiwari}, S. and {Venturi}, T.},
        title = "{Magnetic fields in the Shapley Supercluster core with POSSUM: Challenging model predictions}",
      journal = {\aap},
         year = 2026,
        month = jan,
       volume = {705},
          eid = {A143},
        pages = {A143},
          doi = {10.1051/0004-6361/202556287},
archivePrefix = {arXiv},
       eprint = {2511.14377},
 primaryClass = {astro-ph.CO},
       adsurl = {https://ui.adsabs.harvard.edu/abs/2026A&A...705A.143A}
}

@ARTICLE{anderson2021,
       author = {{Anderson}, C.~S. and {Heald}, G.~H. and {Eilek}, J.~A. and {Lenc}, E. and {Gaensler}, B.~M. and {Rudnick}, Lawrence and {Van Eck}, C.~L. and {O'Sullivan}, S.~P. and {Stil}, J.~M. and {Chippendale}, A. and {Riseley}, C.~J. and {Carretti}, E. and {West}, J. and {Farnes}, J. and {Harvey-Smith}, L. and {McClure-Griffiths}, N.~M. and {Bock}, Douglas C.~J. and {Bunton}, J.~D. and {Koribalski}, B. and {Tremblay}, C.~D. and {Voronkov}, M.~A. and {Warhurst}, K.},
        title = "{Early Science from POSSUM: Shocks, turbulence, and a massive new reservoir of ionised gas in the Fornax cluster}",
      journal = {\pasa},
         year = 2021,
        month = apr,
       volume = {38},
          eid = {e020},
        pages = {e020},
          doi = {10.1017/pasa.2021.4},
archivePrefix = {arXiv},
       eprint = {2102.01702},
 primaryClass = {astro-ph.GA},
       adsurl = {https://ui.adsabs.harvard.edu/abs/2021PASA...38...20A}
}

@ARTICLE{balboni2025A&A...695A.180B,
       author = {{Balboni}, M. and {Ettori}, S. and {Gastaldello}, F. and {Cassano}, R. and {Bonafede}, A. and {Cuciti}, V. and {Botteon}, A. and {Brunetti}, G. and {Bartalucci}, I. and {Gaspari}, M. and {Gavazzi}, R. and {Ghizzardi}, S. and {Gitti}, M. and {Lovisari}, L. and {Maughan}, B.~J. and {Molendi}, S. and {Pointecouteau}, E. and {Pratt}, G.~W. and {Rasia}, E. and {Riva}, G. and {Rossetti}, M. and {Rottgering}, H. and {Sayers}, J. and {van Weeren}, R.~J.},
        title = "{CHEX-MATE: Scaling relations of radio halo profiles for clusters in the LoTSS DR2 area}",
      journal = {\aap},
         year = 2025,
        month = mar,
       volume = {695},
          eid = {A180},
        pages = {A180},
          doi = {10.1051/0004-6361/202453183},
archivePrefix = {arXiv},
       eprint = {2502.18568},
 primaryClass = {astro-ph.CO},
       adsurl = {https://ui.adsabs.harvard.edu/abs/2025A&A...695A.180B}
}

@ARTICLE{Blakeslee2001MNRAS.327.1004B,
       author = {{Blakeslee}, John P. and {Lucey}, John R. and {Barris}, Brian J. and {Hudson}, Michael J. and {Tonry}, John L.},
        title = "{A synthesis of data from fundamental plane and surface brightness fluctuation surveys}",
      journal = {\mnras},
         year = 2001,
        month = nov,
       volume = {327},
       number = {3},
        pages = {1004-1020},
          doi = {10.1046/j.1365-8711.2001.04800.x},
archivePrefix = {arXiv},
       eprint = {astro-ph/0108194},
 primaryClass = {astro-ph},
       adsurl = {https://ui.adsabs.harvard.edu/abs/2001MNRAS.327.1004B}
}

@ARTICLE{bonafede2010,
       author = {{Bonafede}, A. and {Feretti}, L. and {Murgia}, M. and {Govoni}, F. and {Giovannini}, G. and {Dallacasa}, D. and {Dolag}, K. and {Taylor}, G.~B.},
        title = "{The Coma cluster magnetic field from Faraday rotation measures}",
      journal = {\aap},
         year = 2010,
        month = apr,
       volume = {513},
          eid = {A30},
        pages = {A30},
          doi = {10.1051/0004-6361/200913696},
archivePrefix = {arXiv},
       eprint = {1002.0594},
 primaryClass = {astro-ph.CO},
       adsurl = {https://ui.adsabs.harvard.edu/abs/2010A&A...513A..30B}
}

@ARTICLE{BrentjensdeBruyn,
       author = {{Brentjens}, M.~A. and {de Bruyn}, A.~G.},
        title = "{Faraday rotation measure synthesis}",
      journal = {\aap},
         year = 2005,
        month = oct,
       volume = {441},
       number = {3},
        pages = {1217-1228},
          doi = {10.1051/0004-6361:20052990},
archivePrefix = {arXiv},
       eprint = {astro-ph/0507349},
 primaryClass = {astro-ph},
       adsurl = {https://ui.adsabs.harvard.edu/abs/2005A&A...441.1217B}
}

@ARTICLE{Cavaliere1976A&A....49..137C,
       author = {{Cavaliere}, A. and {Fusco-Femiano}, R.},
        title = "{X-rays from hot plasma in clusters of galaxies.}",
      journal = {\aap},
         year = 1976,
        month = may,
       volume = {49},
        pages = {137-144},
       adsurl = {https://ui.adsabs.harvard.edu/abs/1976A&A....49..137C}
}

@ARTICLE{cova2019A&A...628A..83C,
       author = {{Cova}, F. and {Gastaldello}, F. and {Wik}, D.~R. and {Boschin}, W. and {Botteon}, A. and {Brunetti}, G. and {Buote}, D.~A. and {De Grandi}, S. and {Eckert}, D. and {Ettori}, S. and {Feretti}, L. and {Gaspari}, M. and {Ghizzardi}, S. and {Giovannini}, G. and {Girardi}, M. and {Govoni}, F. and {Molendi}, S. and {Murgia}, M. and {Rossetti}, M. and {Vacca}, V.},
        title = "{A joint XMM-NuSTAR observation of the galaxy cluster Abell 523: Constraints on inverse Compton emission}",
      journal = {\aap},
         year = 2019,
        month = aug,
       volume = {628},
          eid = {A83},
        pages = {A83},
          doi = {10.1051/0004-6361/201834644},
archivePrefix = {arXiv},
       eprint = {1906.07730},
 primaryClass = {astro-ph.HE},
       adsurl = {https://ui.adsabs.harvard.edu/abs/2019A&A...628A..83C}
}

@ARTICLE{Costa1988ApJ...327..544D,
       author = {{da Costa}, L. Nicolaci and {Pellegrini}, P.~S. and {Sargent}, W.~L.~W. and {Tonry}, J. and {Davis}, M. and {Meiksin}, A. and {Latham}, David W. and {Menzies}, J.~W. and {Coulson}, I.~A.},
        title = "{The Southern Sky Redshift Survey}",
      journal = {\apj},
         year = 1988,
        month = apr,
       volume = {327},
        pages = {544},
          doi = {10.1086/166215},
       adsurl = {https://ui.adsabs.harvard.edu/abs/1988ApJ...327..544D}
}

@ARTICLE{derubeis2024A&A...691A..23D,
       author = {{De Rubeis}, E. and {Stuardi}, C. and {Bonafede}, A. and {Vazza}, F. and {van Weeren}, R.~J. and {de Gasperin}, F. and {Br{\"u}ggen}, M.},
        title = "{Magnetic fields in the outskirts of PSZ2 G096.88+24.18 from a depolarization analysis of radio relics}",
      journal = {\aap},
         year = 2024,
        month = nov,
       volume = {691},
          eid = {A23},
        pages = {A23},
          doi = {10.1051/0004-6361/202450892},
archivePrefix = {arXiv},
       eprint = {2408.08603},
 primaryClass = {astro-ph.CO},
       adsurl = {https://ui.adsabs.harvard.edu/abs/2024A&A...691A..23D}
}

@ARTICLE{Drinkwater2001ApJ...548L.139D,
       author = {{Drinkwater}, Michael J. and {Gregg}, Michael D. and {Colless}, Matthew},
        title = "{Substructure and Dynamics of the Fornax Cluster}",
      journal = {\apjl},
         year = 2001,
        month = feb,
       volume = {548},
       number = {2},
        pages = {L139-L142},
          doi = {10.1086/319113},
archivePrefix = {arXiv},
       eprint = {astro-ph/0012415},
 primaryClass = {astro-ph},
       adsurl = {https://ui.adsabs.harvard.edu/abs/2001ApJ...548L.139D}
}

@ARTICLE{paola2019MNRAS.486..623D,
       author = {{Dom{\'\i}nguez-Fern{\'a}ndez}, P. and {Vazza}, F. and {Br{\"u}ggen}, M. and {Brunetti}, G.},
        title = "{Dynamical evolution of magnetic fields in the intracluster medium}",
      journal = {\mnras},
         year = 2019,
        month = jun,
       volume = {486},
       number = {1},
        pages = {623-638},
          doi = {10.1093/mnras/stz877},
archivePrefix = {arXiv},
       eprint = {1903.11052},
 primaryClass = {astro-ph.CO},
       adsurl = {https://ui.adsabs.harvard.edu/abs/2019MNRAS.486..623D}
}

@ARTICLE{paola2024ApJ...977..221D,
       author = {{Dom{\'\i}nguez-Fern{\'a}ndez}, P. and {ZuHone}, J. and {Weinberger}, R. and {Bellomi}, E. and {Hernquist}, L. and {Nulsen}, P. and {Brunetti}, G.},
        title = "{Jet Interaction with Galaxy Cluster Mergers}",
      journal = {\apj},
         year = 2024,
        month = dec,
       volume = {977},
       number = {2},
          eid = {221},
        pages = {221},
          doi = {10.3847/1538-4357/ad9028},
archivePrefix = {arXiv},
       eprint = {2406.19681},
 primaryClass = {astro-ph.HE},
       adsurl = {https://ui.adsabs.harvard.edu/abs/2024ApJ...977..221D}
}

@ARTICLE{Donnert2018SSRv..214..122D,
       author = {{Donnert}, J. and {Vazza}, F. and {Br{\"u}ggen}, M. and {ZuHone}, J.},
        title = "{Magnetic Field Amplification in Galaxy Clusters and Its Simulation}",
      journal = {\ssr},
         year = 2018,
        month = dec,
       volume = {214},
       number = {8},
          eid = {122},
        pages = {122},
          doi = {10.1007/s11214-018-0556-8},
archivePrefix = {arXiv},
       eprint = {1810.09783},
 primaryClass = {astro-ph.CO},
       adsurl = {https://ui.adsabs.harvard.edu/abs/2018SSRv..214..122D}
}

@ARTICLE{Eckert2021Univ....7..142E,
       author = {{Eckert}, Dominique and {Gaspari}, Massimo and {Gastaldello}, Fabio and {Le Brun}, Amandine M.~C. and {O'Sullivan}, Ewan},
        title = "{Feedback from Active Galactic Nuclei in Galaxy Groups}",
      journal = {Universe},
         year = 2021,
        month = may,
       volume = {7},
       number = {5},
          eid = {142},
        pages = {142},
          doi = {10.3390/universe7050142},
archivePrefix = {arXiv},
       eprint = {2106.13259},
 primaryClass = {astro-ph.GA},
       adsurl = {https://ui.adsabs.harvard.edu/abs/2021Univ....7..142E}
}

@ARTICLE{ensslin2003A&A...401..835E,
       author = {{En{\ss}lin}, T.~A. and {Vogt}, C.},
        title = "{The magnetic power spectrum in Faraday rotation screens}",
      journal = {\aap},
         year = 2003,
        month = apr,
       volume = {401},
        pages = {835-848},
          doi = {10.1051/0004-6361:20030172},
archivePrefix = {arXiv},
       eprint = {astro-ph/0302426},
 primaryClass = {astro-ph},
       adsurl = {https://ui.adsabs.harvard.edu/abs/2003A&A...401..835E}
}

@INPROCEEDINGS{Felten1996ASPC...88..271F,
       author = {{Felten}, James E.},
        title = "{Mitigating the Baryon Crisis in Clusters: Can Magnetic Pressure be Important?}",
    booktitle = {Clusters, Lensing, and the Future of the Universe},
         year = 1996,
       editor = {{Trimble}, Virginia and {Reisenegger}, Andreas},
       series = {Astronomical Society of the Pacific Conference Series},
       volume = {88},
        month = jan,
        pages = {271},
       adsurl = {https://ui.adsabs.harvard.edu/abs/1996ASPC...88..271F}
}

@ARTICLE{Feretti1999A&A...344..472F,
       author = {{Feretti}, L. and {Dallacasa}, D. and {Govoni}, F. and {Giovannini}, G. and {Taylor}, G.~B. and {Klein}, U.},
        title = "{The radio galaxies and the magnetic field in Abell 119}",
      journal = {\aap},
         year = 1999,
        month = apr,
       volume = {344},
        pages = {472-482},
          doi = {10.48550/arXiv.astro-ph/9902019},
archivePrefix = {arXiv},
       eprint = {astro-ph/9902019},
 primaryClass = {astro-ph},
       adsurl = {https://ui.adsabs.harvard.edu/abs/1999A&A...344..472F}
}

@ARTICLE{Ghirardini2019A&A...621A..41G,
       author = {{Ghirardini}, V. and {Eckert}, D. and {Ettori}, S. and {Pointecouteau}, E. and {Molendi}, S. and {Gaspari}, M. and {Rossetti}, M. and {De Grandi}, S. and {Roncarelli}, M. and {Bourdin}, H. and {Mazzotta}, P. and {Rasia}, E. and {Vazza}, F.},
        title = "{Universal thermodynamic properties of the intracluster medium over two decades in radius in the X-COP sample}",
      journal = {\aap},
         year = 2019,
        month = jan,
       volume = {621},
          eid = {A41},
        pages = {A41},
          doi = {10.1051/0004-6361/201833325},
archivePrefix = {arXiv},
       eprint = {1805.00042},
 primaryClass = {astro-ph.CO},
       adsurl = {https://ui.adsabs.harvard.edu/abs/2019A&A...621A..41G}
}

@ARTICLE{govoni2006,
       author = {{Govoni}, F. and {Murgia}, M. and {Feretti}, L. and {Giovannini}, G. and {Dolag}, K. and {Taylor}, G.~B.},
        title = "{The intracluster magnetic field power spectrum in Abell 2255}",
      journal = {\aap},
         year = 2006,
        month = dec,
       volume = {460},
       number = {2},
        pages = {425-438},
          doi = {10.1051/0004-6361:20065964},
archivePrefix = {arXiv},
       eprint = {astro-ph/0608433},
 primaryClass = {astro-ph},
       adsurl = {https://ui.adsabs.harvard.edu/abs/2006A&A...460..425G}
}

@ARTICLE{govoni2017,
       author = {{Govoni}, F. and {Murgia}, M. and {Vacca}, V. and {Loi}, F. and {Girardi}, M. and {Gastaldello}, F. and {Giovannini}, G. and {Feretti}, L. and {Paladino}, R. and {Carretti}, E. and {Concu}, R. and {Melis}, A. and {Poppi}, S. and {Valente}, G. and {Bernardi}, G. and {Bonafede}, A. and {Boschin}, W. and {Brienza}, M. and {Clarke}, T.~E. and {Colafrancesco}, S. and {de Gasperin}, F. and {Eckert}, D. and {En{\ss}lin}, T.~A. and {Ferrari}, C. and {Gregorini}, L. and {Johnston-Hollitt}, M. and {Junklewitz}, H. and {Orr{\`u}}, E. and {Parma}, P. and {Perley}, R. and {Rossetti}, M. and {B Taylor}, G. and {Vazza}, F.},
        title = "{Sardinia Radio Telescope observations of Abell 194. The intra-cluster magnetic field power spectrum}",
      journal = {\aap},
         year = 2017,
        month = jul,
       volume = {603},
          eid = {A122},
        pages = {A122},
          doi = {10.1051/0004-6361/201630349},
archivePrefix = {arXiv},
       eprint = {1703.08688},
 primaryClass = {astro-ph.CO},
       adsurl = {https://ui.adsabs.harvard.edu/abs/2017A&A...603A.122G}
}

@ARTICLE{guidetti2008,
       author = {{Guidetti}, D. and {Murgia}, M. and {Govoni}, F. and {Parma}, P. and {Gregorini}, L. and {de Ruiter}, H.~R. and {Cameron}, R.~A. and {Fanti}, R.},
        title = "{The intracluster magnetic field power spectrum in Abell 2382}",
      journal = {\aap},
         year = 2008,
        month = jun,
       volume = {483},
       number = {3},
        pages = {699-713},
          doi = {10.1051/0004-6361:20078576},
archivePrefix = {arXiv},
       eprint = {0709.2652},
 primaryClass = {astro-ph},
       adsurl = {https://ui.adsabs.harvard.edu/abs/2008A&A...483..699G}
}

@ARTICLE{guidetti2010,
       author = {{Guidetti}, D. and {Laing}, R.~A. and {Murgia}, M. and {Govoni}, F. and {Gregorini}, L. and {Parma}, P.},
        title = "{Structure of the magnetoionic medium around the Fanaroff-Riley Class I radio galaxy 3C 449}",
      journal = {\aap},
         year = 2010,
        month = may,
       volume = {514},
          eid = {A50},
        pages = {A50},
          doi = {10.1051/0004-6361/200913872},
archivePrefix = {arXiv},
       eprint = {1002.0811},
 primaryClass = {astro-ph.CO},
       adsurl = {https://ui.adsabs.harvard.edu/abs/2010A&A...514A..50G}
}

@ARTICLE{Jensen2001ApJ...550..503J,
       author = {{Jensen}, Joseph B. and {Tonry}, John L. and {Thompson}, Rodger I. and {Ajhar}, Edward A. and {Lauer}, Tod R. and {Rieke}, Marcia J. and {Postman}, Marc and {Liu}, Michael C.},
        title = "{The Infrared Surface Brightness Fluctuation Hubble Constant}",
      journal = {\apj},
         year = 2001,
        month = apr,
       volume = {550},
       number = {2},
        pages = {503-521},
          doi = {10.1086/319819},
archivePrefix = {arXiv},
       eprint = {astro-ph/0011288},
 primaryClass = {astro-ph},
       adsurl = {https://ui.adsabs.harvard.edu/abs/2001ApJ...550..503J}
}

@ARTICLE{khadir2025arXiv251118532K,
       author = {{Khadir}, Affan and {Osinga}, Erik and {Lee}, Wonki and {McConnell}, David and {Gaensler}, B.~M. and {Stuardi}, Chiara and {Anderson}, Craig and {Carretti}, Ettore and {Akahori}, Takuya and {O'Sullivan}, Shane P. and {Baidoo}, Lerato and {West}, Jennifer and {Van Eck}, Cameron and {Rudnick}, Lawrence and {McClure-Griffiths}, Naomi and {Ki}, Yik and {Ma} and {Alonso-L{\'o}pez}, David and {Gordon-Hall}, Paris},
        title = "{Revealing the magnetization of the intracluster medium of Abell 3581 using background Faraday rotation measures from the POSSUM survey}",
      journal = {arXiv e-prints},
         year = 2025,
        month = nov,
          eid = {arXiv:2511.18532},
        pages = {arXiv:2511.18532},
          doi = {10.48550/arXiv.2511.18532},
archivePrefix = {arXiv},
       eprint = {2511.18532},
 primaryClass = {astro-ph.GA},
       adsurl = {https://ui.adsabs.harvard.edu/abs/2025arXiv251118532K}
}

@ARTICLE{laing2008,
       author = {{Laing}, R.~A. and {Bridle}, A.~H. and {Parma}, P. and {Murgia}, M.},
        title = "{Structures of the magnetoionic media around the Fanaroff-Riley Class I radio galaxies 3C31 and Hydra A}",
      journal = {\mnras},
         year = 2008,
        month = dec,
       volume = {391},
       number = {2},
        pages = {521-549},
          doi = {10.1111/j.1365-2966.2008.13895.x},
archivePrefix = {arXiv},
       eprint = {0809.2411},
 primaryClass = {astro-ph},
       adsurl = {https://ui.adsabs.harvard.edu/abs/2008MNRAS.391..521L}
}

@ARTICLE{Lawler1982ApJ...252...81L,
       author = {{Lawler}, J.~M. and {Dennison}, B.},
        title = "{On intracluster Faraday rotation. II - Statistical analysis.}",
      journal = {\apj},
         year = 1982,
        month = jan,
       volume = {252},
        pages = {81-91},
          doi = {10.1086/159536},
       adsurl = {https://ui.adsabs.harvard.edu/abs/1982ApJ...252...81L}
}

@ARTICLE{Loi2019MNRAS.490.4841L,
       author = {{Loi}, F. and {Murgia}, M. and {Govoni}, F. and {Vacca}, V. and {Bonafede}, A. and {Ferrari}, C. and {Prandoni}, I. and {Feretti}, L. and {Giovannini}, G. and {Li}, H.},
        title = "{Rotation measure synthesis applied to synthetic SKA images of galaxy clusters}",
      journal = {\mnras},
         year = 2019,
        month = dec,
       volume = {490},
       number = {4},
        pages = {4841-4857},
          doi = {10.1093/mnras/stz2699},
archivePrefix = {arXiv},
       eprint = {1910.02084},
 primaryClass = {astro-ph.CO},
       adsurl = {https://ui.adsabs.harvard.edu/abs/2019MNRAS.490.4841L}
}

@ARTICLE{Loi2025A&A...694A.125L,
       author = {{Loi}, F. and {Serra}, P. and {Murgia}, M. and {Govoni}, F. and {Vacca}, V. and {Maccagni}, F. and {Kleiner}, D. and {Kamphuis}, P.},
        title = "{The MeerKAT Fornax Survey: IV. A close look at the cluster physics through the densest rotation measure grid}",
      journal = {\aap},
         year = 2025,
        month = feb,
       volume = {694},
          eid = {A125},
        pages = {A125},
          doi = {10.1051/0004-6361/202451711},
archivePrefix = {arXiv},
       eprint = {2501.05519},
 primaryClass = {astro-ph.CO},
       adsurl = {https://ui.adsabs.harvard.edu/abs/2025A&A...694A.125L}
}

@ARTICLE{muller2021,
       author = {{M{\"u}ller}, Ancla and {Pfrommer}, Christoph and {Ignesti}, Alessandro and {Moretti}, Alessia and {Louren{\c{c}}o}, Ana and {Paladino}, Rosita and {Jaff{\'e}}, Yara and {Gitti}, Myriam and {Venturi}, Tiziana and {Gullieuszik}, Marco and {Poggianti}, Bianca and {Vulcani}, Benedetta and {Biviano}, Andrea and {Adebahr}, Bj{\"o}rn and {Dettmar}, Ralf-J{\"u}rgen},
        title = "{Two striking head-tail galaxies in the galaxy cluster IIZW108: insights into transition to turbulence, magnetic fields, and particle re-acceleration}",
      journal = {\mnras},
         year = 2021,
        month = dec,
       volume = {508},
       number = {4},
        pages = {5326-5344},
          doi = {10.1093/mnras/stab2928},
archivePrefix = {arXiv},
       eprint = {2110.03705},
 primaryClass = {astro-ph.GA},
       adsurl = {https://ui.adsabs.harvard.edu/abs/2021MNRAS.508.5326M}
}

@ARTICLE{murgia2004,
       author = {{Murgia}, M. and {Govoni}, F. and {Feretti}, L. and {Giovannini}, G. and {Dallacasa}, D. and {Fanti}, R. and {Taylor}, G.~B. and {Dolag}, K.},
        title = "{Magnetic fields and Faraday rotation in clusters of galaxies}",
      journal = {\aap},
         year = 2004,
        month = sep,
       volume = {424},
        pages = {429-446},
          doi = {10.1051/0004-6361:20040191},
archivePrefix = {arXiv},
       eprint = {astro-ph/0406225},
 primaryClass = {astro-ph},
       adsurl = {https://ui.adsabs.harvard.edu/abs/2004A&A...424..429M}
}

@ARTICLE{Norseth2025arXiv250815138N,
       author = {{Norseth}, Christian and {Wik}, Daniel R. and {Sarazin}, Craig L. and {Sun}, Ming and {Gastaldello}, Fabio},
        title = "{Electron-Ion Equilibration in the Merging Galaxy Cluster Abell 665}",
      journal = {arXiv e-prints},
         year = 2025,
        month = aug,
          eid = {arXiv:2508.15138},
        pages = {arXiv:2508.15138},
          doi = {10.48550/arXiv.2508.15138},
archivePrefix = {arXiv},
       eprint = {2508.15138},
 primaryClass = {astro-ph.CO},
       adsurl = {https://ui.adsabs.harvard.edu/abs/2025arXiv250815138N}
}

@ARTICLE{Osinga2025A&A...694A..44O,
       author = {{Osinga}, E. and {van Weeren}, R.~J. and {Rudnick}, L. and {Andrade-Santos}, F. and {Bonafede}, A. and {Clarke}, T. and {Duncan}, K. and {Giacintucci}, S. and {R{\"o}ttgering}, H.~J.~A.},
        title = "{Probing cluster magnetism with embedded and background radio sources in Planck clusters}",
      journal = {\aap},
         year = 2025,
        month = feb,
       volume = {694},
          eid = {A44},
        pages = {A44},
          doi = {10.1051/0004-6361/202451885},
archivePrefix = {arXiv},
       eprint = {2408.07178},
 primaryClass = {astro-ph.CO},
       adsurl = {https://ui.adsabs.harvard.edu/abs/2025A&A...694A..44O}
}

@ARTICLE{Pagliotta2025A&A...700A.139P,
       author = {{Pagliotta}, A. and {Riseley}, C.~J. and {Bonafede}, A. and {Stuardi}, C. and {Loi}, F.},
        title = "{Constraining the magnetic field in the galaxy cluster Abell 2142 using MeerKAT L-band polarisation data}",
      journal = {\aap},
         year = 2025,
        month = aug,
       volume = {700},
          eid = {A139},
        pages = {A139},
          doi = {10.1051/0004-6361/202554881},
archivePrefix = {arXiv},
       eprint = {2507.00114},
 primaryClass = {astro-ph.CO},
       adsurl = {https://ui.adsabs.harvard.edu/abs/2025A&A...700A.139P}
}

@INPROCEEDINGS{Pfrommer2017AIPC.1792c0003P,
       author = {{Pfrommer}, C. and {Pakmor}, R. and {Schaal}, K. and {Simpson}, C.~M. and {Springel}, V.},
        title = "{Cosmic ray feedback in galaxies and active galactic nuclei}",
    booktitle = {6th International Symposium on High Energy Gamma-Ray Astronomy},
         year = 2017,
       series = {American Institute of Physics Conference Series},
       volume = {1792},
        month = jan,
          eid = {030003},
        pages = {030003},
          doi = {10.1063/1.4968904},
       adsurl = {https://ui.adsabs.harvard.edu/abs/2017AIPC.1792c0003P}
}

@ARTICLE{piffaretti2011,
       author = {{Piffaretti}, R. and {Arnaud}, M. and {Pratt}, G.~W. and {Pointecouteau}, E. and {Melin}, J. -B.},
        title = "{The MCXC: a meta-catalogue of x-ray detected clusters of galaxies}",
      journal = {\aap},
         year = 2011,
        month = oct,
       volume = {534},
          eid = {A109},
        pages = {A109},
          doi = {10.1051/0004-6361/201015377},
archivePrefix = {arXiv},
       eprint = {1007.1916},
 primaryClass = {astro-ph.CO},
       adsurl = {https://ui.adsabs.harvard.edu/abs/2011A&A...534A.109P}
}

@ARTICLE{raj2024arXiv240703225R,
       author = {{Raj}, Maria Angela and {Awad}, Petra and {Peletier}, Reynier F. and {Smith}, Rory and {Kuchner}, Ulrike and {van de Weygaert}, Rien and {Libeskind}, Noam I. and {Canducci}, Marco and {Tino}, Peter and {Bunte}, Kerstin},
        title = "{The large-scale structure around the Fornax-Eridanus Complex}",
      journal = {arXiv e-prints},
         year = 2024,
        month = jul,
          eid = {arXiv:2407.03225},
        pages = {arXiv:2407.03225},
          doi = {10.48550/arXiv.2407.03225},
archivePrefix = {arXiv},
       eprint = {2407.03225},
 primaryClass = {astro-ph.GA},
       adsurl = {https://ui.adsabs.harvard.edu/abs/2024arXiv240703225R}
}

@ARTICLE{Rappaz2024A&A...691A.132R,
       author = {{Rappaz}, Y. and {Schober}, J. and {Bendre}, A.~B. and {Seta}, A. and {Federrath}, C.},
        title = "{Cosmic evolution of the Faraday rotation measure in the intracluster medium of galaxy clusters}",
      journal = {\aap},
         year = 2024,
        month = nov,
       volume = {691},
          eid = {A132},
        pages = {A132},
          doi = {10.1051/0004-6361/202451119},
archivePrefix = {arXiv},
       eprint = {2409.18580},
 primaryClass = {astro-ph.CO},
       adsurl = {https://ui.adsabs.harvard.edu/abs/2024A&A...691A.132R}
}

@ARTICLE{Reiprich2002ApJ...567..716R,
       author = {{Reiprich}, Thomas H. and {B{\"o}hringer}, Hans},
        title = "{The Mass Function of an X-Ray Flux-limited Sample of Galaxy Clusters}",
      journal = {\apj},
         year = 2002,
        month = mar,
       volume = {567},
       number = {2},
        pages = {716-740},
          doi = {10.1086/338753},
archivePrefix = {arXiv},
       eprint = {astro-ph/0111285},
 primaryClass = {astro-ph},
       adsurl = {https://ui.adsabs.harvard.edu/abs/2002ApJ...567..716R}
}

@ARTICLE{Reiprich2025arXiv250302884R,
       author = {{Reiprich}, T.~H. and {Veronica}, A. and {Pacaud}, F. and {St{\"o}cker}, P. and {Nazaretyan}, V. and {Srivastava}, A. and {Pandya}, A. and {Dietl}, J. and {Sanders}, J.~S. and {Yeung}, M.~C.~H. and {Chaturvedi}, A. and {Hilker}, M. and {Seidel}, B. and {Dolag}, K. and {Comparat}, J. and {Ghirardini}, V. and {Kluge}, M. and {Liu}, A. and {Malavasi}, N. and {Zhang}, X. and {Hern{\'a}ndez-Mart{\'\i}nez}, E.},
        title = "{The SRG/eROSITA all-sky survey: View of the Fornax galaxy cluster}",
      journal = {arXiv e-prints},
         year = 2025,
        month = mar,
          eid = {arXiv:2503.02884},
        pages = {arXiv:2503.02884},
          doi = {10.48550/arXiv.2503.02884},
archivePrefix = {arXiv},
       eprint = {2503.02884},
 primaryClass = {astro-ph.CO},
       adsurl = {https://ui.adsabs.harvard.edu/abs/2025arXiv250302884R}
}

@ARTICLE{Ruszkowski2010ApJ...713.1332R,
       author = {{Ruszkowski}, M. and {Oh}, S. Peng},
        title = "{Shaken and Stirred: Conduction and Turbulence in Clusters of Galaxies}",
      journal = {\apj},
         year = 2010,
        month = apr,
       volume = {713},
       number = {2},
        pages = {1332-1342},
          doi = {10.1088/0004-637X/713/2/1332},
archivePrefix = {arXiv},
       eprint = {0911.5198},
 primaryClass = {astro-ph.CO},
       adsurl = {https://ui.adsabs.harvard.edu/abs/2010ApJ...713.1332R}
}

@ARTICLE{Simionescu2019SSRv..215...24S,
       author = {{Simionescu}, Aurora and {ZuHone}, John and {Zhuravleva}, Irina and {Churazov}, Eugene and {Gaspari}, Massimo and {Nagai}, Daisuke and {Werner}, Norbert and {Roediger}, Elke and {Canning}, Rebecca and {Eckert}, Dominique and {Gu}, Liyi and {Paerels}, Frits},
        title = "{Constraining Gas Motions in the Intra-Cluster Medium}",
      journal = {\ssr},
         year = 2019,
        month = feb,
       volume = {215},
       number = {2},
          eid = {24},
        pages = {24},
          doi = {10.1007/s11214-019-0590-1},
archivePrefix = {arXiv},
       eprint = {1902.00024},
 primaryClass = {astro-ph.CO},
       adsurl = {https://ui.adsabs.harvard.edu/abs/2019SSRv..215...24S}
}

@ARTICLE{Sotomayor2013A&A...552A..58S,
       author = {{Sotomayor-Beltran}, C. and {Sobey}, C. and {Hessels}, J.~W.~T. and {de Bruyn}, G. and {Noutsos}, A. and {Alexov}, A. and {Anderson}, J. and {Asgekar}, A. and {Avruch}, I.~M. and {Beck}, R. and {Bell}, M.~E. and {Bell}, M.~R. and {Bentum}, M.~J. and {Bernardi}, G. and {Best}, P. and {Birzan}, L. and {Bonafede}, A. and {Breitling}, F. and {Broderick}, J. and {Brouw}, W.~N. and {Br{\"u}ggen}, M. and {Ciardi}, B. and {de Gasperin}, F. and {Dettmar}, R. -J. and {van Duin}, A. and {Duscha}, S. and {Eisl{\"o}ffel}, J. and {Falcke}, H. and {Fallows}, R.~A. and {Fender}, R. and {Ferrari}, C. and {Frieswijk}, W. and {Garrett}, M.~A. and {Grie{\ss}meier}, J. and {Grit}, T. and {Gunst}, A.~W. and {Hassall}, T.~E. and {Heald}, G. and {Hoeft}, M. and {Horneffer}, A. and {Iacobelli}, M. and {Juette}, E. and {Karastergiou}, A. and {Keane}, E. and {Kohler}, J. and {Kramer}, M. and {Kondratiev}, V.~I. and {Koopmans}, L.~V.~E. and {Kuniyoshi}, M. and {Kuper}, G. and {van Leeuwen}, J. and {Maat}, P. and {Macario}, G. and {Markoff}, S. and {McKean}, J.~P. and {Mulcahy}, D.~D. and {Munk}, H. and {Orru}, E. and {Paas}, H. and {Pandey-Pommier}, M. and {Pilia}, M. and {Pizzo}, R. and {Polatidis}, A.~G. and {Reich}, W. and {R{\"o}ttgering}, H. and {Serylak}, M. and {Sluman}, J. and {Stappers}, B.~W. and {Tagger}, M. and {Tang}, Y. and {Tasse}, C. and {ter Veen}, S. and {Vermeulen}, R. and {van Weeren}, R.~J. and {Wijers}, R.~A.~M.~J. and {Wijnholds}, S.~J. and {Wise}, M.~W. and {Wucknitz}, O. and {Yatawatta}, S. and {Zarka}, P.},
        title = "{Calibrating high-precision Faraday rotation measurements for LOFAR and the next generation of low-frequency radio telescopes}",
      journal = {\aap},
         year = 2013,
        month = apr,
       volume = {552},
          eid = {A58},
        pages = {A58},
          doi = {10.1051/0004-6361/201220728},
archivePrefix = {arXiv},
       eprint = {1303.6230},
 primaryClass = {astro-ph.IM},
       adsurl = {https://ui.adsabs.harvard.edu/abs/2013A&A...552A..58S}
}

@ARTICLE{stuardi2021,
       author = {{Stuardi}, C. and {Bonafede}, A. and {Lovisari}, L. and {Dom{\'\i}nguez-Fern{\'a}ndez}, P. and {Vazza}, F. and {Br{\"u}ggen}, M. and {van Weeren}, R.~J. and {de Gasperin}, F.},
        title = "{The intracluster magnetic field in the double relic galaxy cluster Abell 2345}",
      journal = {\mnras},
         year = 2021,
        month = apr,
       volume = {502},
       number = {2},
        pages = {2518-2535},
          doi = {10.1093/mnras/stab218},
archivePrefix = {arXiv},
       eprint = {2101.09302},
 primaryClass = {astro-ph.GA},
       adsurl = {https://ui.adsabs.harvard.edu/abs/2021MNRAS.502.2518S}
}

@ARTICLE{su2017ApJ...851...69S,
       author = {{Su}, Yuanyuan and {Nulsen}, Paul E.~J. and {Kraft}, Ralph P. and {Roediger}, Elke and {ZuHone}, John A. and {Jones}, Christine and {Forman}, William R. and {Sheardown}, Alex and {Irwin}, Jimmy A. and {Randall}, Scott W.},
        title = "{Gas Sloshing Regulates and Records the Evolution of the Fornax Cluster}",
      journal = {\apj},
         year = 2017,
        month = dec,
       volume = {851},
       number = {1},
          eid = {69},
        pages = {69},
          doi = {10.3847/1538-4357/aa989e},
archivePrefix = {arXiv},
       eprint = {1711.01523},
 primaryClass = {astro-ph.HE},
       adsurl = {https://ui.adsabs.harvard.edu/abs/2017ApJ...851...69S}
}

@ARTICLE{Subramanian2006MNRAS.366.1437S,
       author = {{Subramanian}, Kandaswamy and {Shukurov}, Anvar and {Haugen}, Nils Erland L.},
        title = "{Evolving turbulence and magnetic fields in galaxy clusters}",
      journal = {\mnras},
         year = 2006,
        month = mar,
       volume = {366},
       number = {4},
        pages = {1437-1454},
          doi = {10.1111/j.1365-2966.2006.09918.x},
archivePrefix = {arXiv},
       eprint = {astro-ph/0505144},
 primaryClass = {astro-ph},
       adsurl = {https://ui.adsabs.harvard.edu/abs/2006MNRAS.366.1437S}
}

@ARTICLE{Tevlin2025A&A...701A.114T,
       author = {{Tevlin}, L. and {Berlok}, T. and {Pfrommer}, C. and {Talbot}, R.~Y. and {Whittingham}, J. and {Puchwein}, E. and {Pakmor}, R. and {Weinberger}, R. and {Springel}, V.},
        title = "{Magnetic dynamos in galaxy clusters: The crucial role of galaxy formation physics at high redshifts}",
      journal = {\aap},
         year = 2025,
        month = sep,
       volume = {701},
          eid = {A114},
        pages = {A114},
          doi = {10.1051/0004-6361/202452823},
archivePrefix = {arXiv},
       eprint = {2411.00103},
 primaryClass = {astro-ph.GA},
       adsurl = {https://ui.adsabs.harvard.edu/abs/2025A&A...701A.114T}
}

@ARTICLE{Tonry2001ApJ...546..681T,
       author = {{Tonry}, John L. and {Dressler}, Alan and {Blakeslee}, John P. and {Ajhar}, Edward A. and {Fletcher}, Andr{\'e} B. and {Luppino}, Gerard A. and {Metzger}, Mark R. and {Moore}, Christopher B.},
        title = "{The SBF Survey of Galaxy Distances. IV. SBF Magnitudes, Colors, and Distances}",
      journal = {\apj},
         year = 2001,
        month = jan,
       volume = {546},
       number = {2},
        pages = {681-693},
          doi = {10.1086/318301},
archivePrefix = {arXiv},
       eprint = {astro-ph/0011223},
 primaryClass = {astro-ph},
       adsurl = {https://ui.adsabs.harvard.edu/abs/2001ApJ...546..681T}
}

@ARTICLE{vacca2012,
       author = {{Vacca}, V. and {Murgia}, M. and {Govoni}, F. and {Feretti}, L. and {Giovannini}, G. and {Perley}, R.~A. and {Taylor}, G.~B.},
        title = "{The intracluster magnetic field power spectrum in A2199}",
      journal = {\aap},
         year = 2012,
        month = apr,
       volume = {540},
          eid = {A38},
        pages = {A38},
          doi = {10.1051/0004-6361/201116622},
archivePrefix = {arXiv},
       eprint = {1201.4119},
 primaryClass = {astro-ph.CO},
       adsurl = {https://ui.adsabs.harvard.edu/abs/2012A&A...540A..38V}
}

@ARTICLE{Vacca2022MNRAS.514.4969V,
       author = {{Vacca}, Valentina and {Govoni}, Federica and {Murgia}, Matteo and {Perley}, Richard A. and {Feretti}, Luigina and {Giovannini}, Gabriele and {Carretti}, Ettore and {Gastaldello}, Fabio and {Cova}, Filippo and {Marchegiani}, Paolo and {Battistelli}, Elia and {Boschin}, Walter and {En{\ss}lin}, Torsten A. and {Girardi}, Marisa and {Loi}, Francesca and {Radiconi}, Federico},
        title = "{Puzzling large-scale polarization in the galaxy cluster Abell 523}",
      journal = {\mnras},
         year = 2022,
        month = aug,
       volume = {514},
       number = {4},
        pages = {4969-4981},
          doi = {10.1093/mnras/stac1421},
archivePrefix = {arXiv},
       eprint = {2206.03402},
 primaryClass = {astro-ph.CO},
       adsurl = {https://ui.adsabs.harvard.edu/abs/2022MNRAS.514.4969V}
}

@ARTICLE{vanweeren2019,
       author = {{van Weeren}, R.~J. and {de Gasperin}, F. and {Akamatsu}, H. and {Br{\"u}ggen}, M. and {Feretti}, L. and {Kang}, H. and {Stroe}, A. and {Zandanel}, F.},
        title = "{Diffuse Radio Emission from Galaxy Clusters}",
      journal = {\ssr},
         year = 2019,
        month = feb,
       volume = {215},
       number = {1},
          eid = {16},
        pages = {16},
          doi = {10.1007/s11214-019-0584-z},
archivePrefix = {arXiv},
       eprint = {1901.04496},
 primaryClass = {astro-ph.HE},
       adsurl = {https://ui.adsabs.harvard.edu/abs/2019SSRv..215...16V}
}

@ARTICLE{venturi2022A&A...660A..81V,
       author = {{Venturi}, T. and {Giacintucci}, S. and {Merluzzi}, P. and {Bardelli}, S. and {Busarello}, G. and {Dallacasa}, D. and {Sikhosana}, S.~P. and {Marvil}, J. and {Smirnov}, O. and {Bourdin}, H. and {Mazzotta}, P. and {Rossetti}, M. and {Rudnick}, L. and {Bernardi}, G. and {Br{\"u}ggen}, M. and {Carretti}, E. and {Cassano}, R. and {Di Gennaro}, G. and {Gastaldello}, F. and {Kale}, R. and {Knowles}, K. and {Koribalski}, B.~S. and {Heywood}, I. and {Hopkins}, A.~M. and {Norris}, R.~P. and {Reiprich}, T.~H. and {Tasse}, C. and {Vernstrom}, T. and {Zucca}, E. and {Bester}, L.~H. and {Diego}, J.~M. and {Kanapathippillai}, J.},
        title = "{Radio footprints of a minor merger in the Shapley Supercluster: From supercluster down to galactic scales}",
      journal = {\aap},
         year = 2022,
        month = apr,
       volume = {660},
          eid = {A81},
        pages = {A81},
          doi = {10.1051/0004-6361/202142048},
archivePrefix = {arXiv},
       eprint = {2201.04887},
 primaryClass = {astro-ph.CO},
       adsurl = {https://ui.adsabs.harvard.edu/abs/2022A&A...660A..81V}
}

@ARTICLE{Schekochihin2004ApJ...612..276S,
       author = {{Schekochihin}, Alexander A. and {Cowley}, Steven C. and {Taylor}, Samuel F. and {Maron}, Jason L. and {McWilliams}, James C.},
        title = "{Simulations of the Small-Scale Turbulent Dynamo}",
      journal = {\apj},
         year = 2004,
        month = sep,
       volume = {612},
       number = {1},
        pages = {276-307},
          doi = {10.1086/422547},
archivePrefix = {arXiv},
       eprint = {astro-ph/0312046},
 primaryClass = {astro-ph},
       adsurl = {https://ui.adsabs.harvard.edu/abs/2004ApJ...612..276S}
}

@ARTICLE{serra2023,
       author = {{Serra}, P. and {Maccagni}, F.~M. and {Kleiner}, D. and {Moln{\'a}r}, D. and {Ramatsoku}, M. and {Loni}, A. and {Loi}, F. and {de Blok}, W.~J.~G. and {Bryan}, G.~L. and {Dettmar}, R.~J. and {Frank}, B.~S. and {van Gorkom}, J.~H. and {Govoni}, F. and {Iodice}, E. and {J{\'o}zsa}, G.~I.~G. and {Kamphuis}, P. and {Kraan-Korteweg}, R. and {Loubser}, S.~I. and {Murgia}, M. and {Oosterloo}, T.~A. and {Peletier}, R. and {Pisano}, D.~J. and {Smith}, M.~W.~L. and {Trager}, S.~C. and {Verheijen}, M.~A.~W.},
        title = "{The MeerKAT Fornax Survey. I. Survey description and first evidence of ram pressure in the Fornax galaxy cluster}",
      journal = {\aap},
         year = 2023,
        month = may,
       volume = {673},
          eid = {A146},
        pages = {A146},
          doi = {10.1051/0004-6361/202346071},
archivePrefix = {arXiv},
       eprint = {2302.11895},
 primaryClass = {astro-ph.GA},
       adsurl = {https://ui.adsabs.harvard.edu/abs/2023A&A...673A.146S}
}

@ARTICLE{vacca2010A&A...514A..71V,
       author = {{Vacca}, V. and {Murgia}, M. and {Govoni}, F. and {Feretti}, L. and {Giovannini}, G. and {Orr{\`u}}, E. and {Bonafede}, A.},
        title = "{The intracluster magnetic field power spectrum in Abell 665}",
      journal = {\aap},
         year = 2010,
        month = may,
       volume = {514},
          eid = {A71},
        pages = {A71},
          doi = {10.1051/0004-6361/200913060},
       adsurl = {https://ui.adsabs.harvard.edu/abs/2010A&A...514A..71V}
}

@ARTICLE{vogt2003A&A...412..373V,
       author = {{Vogt}, C. and {En{\ss}lin}, T.~A.},
        title = "{Measuring the cluster magnetic field power spectra from Faraday rotation maps of Abell 400, Abell 2634 and Hydra A}",
      journal = {\aap},
         year = 2003,
        month = dec,
       volume = {412},
        pages = {373-385},
          doi = {10.1051/0004-6361:20031434},
archivePrefix = {arXiv},
       eprint = {astro-ph/0309441},
 primaryClass = {astro-ph},
       adsurl = {https://ui.adsabs.harvard.edu/abs/2003A&A...412..373V}
}

@ARTICLE{Watson2023ApJ...955..103W,
       author = {{Watson}, Courtney B. and {Blanton}, Elizabeth L. and {Randall}, Scott W. and {Sarazin}, Craig L. and {Sarkar}, Arnab and {ZuHone}, John A. and {Douglass}, E.~M.},
        title = "{CHANDRA X-Ray Observations of A119: Cold Fronts and a Shock in an Evolved Off-axis Merger}",
      journal = {\apj},
         year = 2023,
        month = oct,
       volume = {955},
       number = {2},
          eid = {103},
        pages = {103},
          doi = {10.3847/1538-4357/acee74},
archivePrefix = {arXiv},
       eprint = {2308.04367},
 primaryClass = {astro-ph.GA},
       adsurl = {https://ui.adsabs.harvard.edu/abs/2023ApJ...955..103W}
}

@ARTICLE{Zhang2022MNRAS.516...26Z,
       author = {{Zhang}, Bowei and {Cui}, Weiguang and {Wang}, Yuhuan and {Dave}, Romeel and {De Petris}, Marco},
        title = "{THE THREE HUNDRED: cluster dynamical states and relaxation period}",
      journal = {\mnras},
         year = 2022,
        month = oct,
       volume = {516},
       number = {1},
        pages = {26-38},
          doi = {10.1093/mnras/stac2171},
archivePrefix = {arXiv},
       eprint = {2112.01909},
 primaryClass = {astro-ph.CO},
       adsurl = {https://ui.adsabs.harvard.edu/abs/2022MNRAS.516...26Z}
}

@ARTICLE{ZuHone2011ApJ...743...16Z,
       author = {{ZuHone}, J.~A. and {Markevitch}, M. and {Lee}, D.},
        title = "{Sloshing of the Magnetized Cool Gas in the Cores of Galaxy Clusters}",
      journal = {\apj},
         year = 2011,
        month = dec,
       volume = {743},
       number = {1},
          eid = {16},
        pages = {16},
          doi = {10.1088/0004-637X/743/1/16},
archivePrefix = {arXiv},
       eprint = {1108.4427},
 primaryClass = {astro-ph.CO},
       adsurl = {https://ui.adsabs.harvard.edu/abs/2011ApJ...743...16Z}
}

\begin{appendix}
\section{List of magnetic field measurements}
\begin{table*}
\caption{List of magnetic field measurements in galaxy clusters and groups derived from spectro-polarimetric modelling of cluster diffuse sources or RM analysis.}
\label{tab:literature}
\begin{tabular}{lllllllllllllll}
\hline
\hline
 Name & z  & type & dens. & n$_0$ & r$_c$ & $\beta$ & P.S. & $\zeta$  & $\Lambda_{min}$ & $\Lambda_{max}$ & B$_0$ & $\eta$ & M$_{500}$  &   R$_{500}$  \\
      &    &      & mod. & \small{10$^{-3}$} & \small{kpc} &  & mod. &  &  \small{kpc}  & \small{kpc} &      \small{$\muup$G} &   &  \small{10$^{14}$M$_{\odot}$}  & \small{Mpc}  \\
\hline
\hline
 3C31           & 0.0169   & GG     & $\beta$         & 1.9   & 51     & 0.38    & bpl       & 2.32  & 6      & 452      & 6.7  & 1     & 0.4 & 0.52 \\
                &          &        &                 &       &        &         &              & 3.$\overline{6}$ & 5.5    &           &      &       &                    &      \\ 
 3C449          & 0.017    & GG     & $\beta$         & 3.7   & 57.1   & 0.42    & bpl          & 2.07  & 0.2*    & 65      & 3.5  & 1     & &        \\
                &          &        &                 &       &        &         &              & 2.98  & 11     &          &      &       &              &        \\
A119           & 0.044    & MMC    & $\beta$         & 1.4   & 379    & 0.56    & pl           & 2     & 6      & 772      & 6    & 0.89  & 2.5 & 0.94  \\
 A194           & 0.0179   & LRC    & $\beta$         & 0.69  & 245  & 0.67    & pl           & 3.$\overline{6}$* & 1*      & 64      & 1.5  & 1.1   & 0.4 & 0.52  \\
 A2142          & 0.0894   & MMC    & univ.           & 7.44  & 0.07 & 0.39    & df           &       & 7*      & 463     & 6    & 0.5*   & 8.1  & 1.38 \\
                &          &        &                 & 0.7   & 0.6    & 2.6     &              & 1.1*   & 0.005*  &        &      &       &          &        \\
 A2199          & 0.0311   & MRC    & 2$\beta$        & 74    & 9      & 1.5     & pl           & 2.8   & 0.7    & 35       & 11.7 & 0.9   & 2.9 & 1.0 \\
                &          &        &                 & 27    & 26     & 0.39    &              &       &        &          &      &       &             &        \\
 A2255          & 0.0806   & MMC    & $\beta$         & 2.05  & 432    & 0.74    & pl           & 3     & 4*      & 512*    & 2    & 0.5*   & 3.7  & 1.07 \\
 A2345          & 0.1789   & MMC    & d$\beta$        & 2.8   & 192    & 0.67    & df           &       & 8*      & 978*    & 2.8  & 1     & 4.1  & 1.07 \\
                &          &        &                 & 0.56  & 2148   & 1.48    &              & 1.054* & 0.004*  &        &      &       &               &        \\
 A2382          & 0.0618   & MRC    & 2$\beta$        & 1.2   & 374  & 0.9     & pl           & 3.$\overline{6}$ & 12*     & 70       & 3    & 0.5   & 1.1  & 0.71 \\
                &          &        &                 & 3.8   & 65.7   & 1.7     &              &       &        &          &      &       &          &        \\
 A2634          & 0.0302   & MRC    & $\beta$         & 1.2   & 178    & 0.79    & pl           & 3.6   & 2.2    & 10.5     & 3    & 1     & 1.2 & 0.75  \\
 A3581          & 0.0221   & MRC    & $\beta$         & 33.6  & 10.2  & 0.47    & pl           & 3.$\overline{6}$* & 4*      & 100*    & 2.2  & 0.5   & 1.1 & 0.72 \\
 A400           & 0.0231  & LMC    & $\beta$         & 1.6   & 109    & 0.54    & pl           & 3.8   & 1.4    & 7.8       & 6    & 1     & 0.8  & 0.65  \\
 A523           & 0.104    & MMC    & $\beta$         & 1.1   & 185    & 0.42    & bpl          & 3.$\overline{6}$* & 7.9*    & 2013*   & 0.5  & 0.5   & 1.8 & 0.83  \\
                &          &        &                 &       &        &         &              & 0*     & 1006*   &        &      &       &               &        \\ A665           & 0.1829   & MMC    & $\beta$         & 3.25  & 340 & 0.763   & pl           & 3.$\overline{6}$* & 4*      & 450     & 1.3  & 0.47  & 6.9  & 1.26 \\
 Coma       & 0.023    & MMC    & $\beta$         & 3.44  & 291    & 0.75    & pl           & 3.$\overline{6}$* & 2      & 34       & 4.7  & 0.5   & 4.3   & 1.14 \\
 Hydra A        & 0.0549   & MRC    & 2$\beta$        & 62    & 27     & 0.686   & pl        & 2.77  & 0.62   & 87       & 45.5 & 0.25  & 3.6  & 1.07 \\
                &          &        &                 & 0.26 & 232    & 0.907   &              &       &        &          &      &       &            &        \\
 Fornax     & 0.0049   & LRC      & $\beta$         & 1.07 & 130    & 0.7     & pl           & 2.7   & 1      & 15       & 5    & 1.6   & 0.2 & 0.4 \\
\hline
\end{tabular}
\tablefoot{Columns from left to right report: the system name; redshift; classification (GG: Galaxy Group, MMC: Massive Merging Cluster, MRC: Massive Relaxed Cluster, LMC: Low-mass Merging Cluster, LRC: Low-mass Relaxed Cluster); the assumed density model ($\beta$ for a single $\beta$-model, 2$\beta$ for a sum of two $\beta$-models, d$\beta$ for a sum in quadrature, and univ. for the universal profile introduced by \citealt{Ghirardini2019A&A...621A..41G}); central thermal density $n_0$; core radius $r_c$; and the $\beta$ index. Additional parameters are provided in the subsequent row where necessary; specifically, for the universal model, the second $n_0$, $\beta$, and $r_c$ values correspond to $\alpha$, $\epsilon$, and $R_s$, respectively. The table also lists: the magnetic field power spectrum model ('pl' for power law, 'bpl' for broken power law, and 'df' for the model proposed by \citealt{paola2019MNRAS.486..623D}); the spectral slope $\zeta$; and the $\Lambda_{\rm min}$ and $\Lambda_{\rm max}$ scales. For the 'df' profile, $\zeta$ and the second $\Lambda_{\rm min}$ correspond to parameters $B$ and $C$, respectively. Radial profile parameters for the magnetic field ($B_0$ and $\eta$) are also included. We note that 3C449 is the only case with a broken profile, characterized by $\eta=0$ at radial distances exceeding 16\,kpc. Finally, we report the mass and radius computed within $R_{500}$ \citep{piffaretti2011}. Parameters assumed a priori in the original modelling are highlighted with an asterisk. All spatial scales, densities, and magnetic field strengths have been rescaled to the cosmology adopted in this work. For 3C31 and Hydra A, we rescaled the reported $B_{\rm rms}$ values according to $B_{\rm rms} = 0.42 B_0$ \citep{laing2008}. For A2255, we refer to the modelling based on radio galaxy RMs \citep{govoni2006}. In the case of 3C31, we used results obtained without cavity modelling, as these features were not detected. For both A2345 and A2142, $B_0$ was extrapolated from the average magnetic field strength within a 1\,Mpc$^3$ volume reported by the authors, assuming the radial profile parameters ($\eta$, $n_e$) derived for the system. Central temperatures for A665 and A523 were taken from \citet{Norseth2025arXiv250815138N} and \citet{cova2019A&A...628A..83C}, respectively. 
Literature data are sourced from: 
\cite{guidetti2010} (3C449), \cite{laing2008} (3C31, Hydra A), \cite{murgia2004} (A119), \cite{govoni2017} (A194),\cite{Pagliotta2025A&A...700A.139P} (A2142), \cite{vacca2012} (A2199), \cite{govoni2006} (A2255), \cite{stuardi2021} (A2345), \cite{guidetti2008} (A2382), \cite{vogt2003A&A...412..373V} (A2634 and A400), \cite{khadir2025arXiv251118532K} (A3581), \cite{Vacca2022MNRAS.514.4969V} (A523), \cite{vacca2010A&A...514A..71V} (A665), \cite{bonafede2010} (Coma Cluster), and this work (Fornax cluster).}
\end{table*}
\FloatBarrier
\end{appendix}
\end{document}